\documentclass[onecolumn]{aastex6}

\usepackage{color,txfonts}
\usepackage{hyperref}
\usepackage{epstopdf}
\usepackage{graphicx}
\usepackage[FIGTOPCAP]{subfigure}

\begin{document}

\title{GW170817 and the prospect of forming supramassive remnants in neutron star mergers}

\author{Peng-Xiong Ma$^{1,2}$, Jin-Liang Jiang$^{1,2}$, Hao Wang$^{1,3}$, Zhi-Ping Jin$^{1,2}$,
 Yi-Zhong Fan$^{1,2}$, and Da-Ming Wei$^{1,2}$}
\affil{
$^1$ {Key Laboratory of dark Matter and Space Astronomy, Purple Mountain Observatory, Chinese Academy of Science, Nanjing, 210008, China.}\\
$^2$ {School of Astronomy and Space Science, University of Science and Technology of China, Hefei, Anhui 230026, China.}\\
$^3$ {University of Chinese Academy of Sciences, Yuquan Road 19, Beijing, 100049, China}\\
}
\email{yzfan@pmo.ac.cn (YZF)}

\begin{abstract}
The gravitational wave data of GW170817 favor the equation of state (EoS) models that predict compact neutron stars (NSs), consistent with the
radius constraints from X-ray observations. Motivated by such a remarkable progress,
we examine the fate of the remnants formed in NS mergers and focus on the roles of the angular momentum and the
mass distribution of the binary NSs. In the mass shedding limit (for which the dimensionless angular momentum equals to the Keplerian value, i.e., $j=j_{\rm Kep}$), the adopted { seven EoS models, except H4 and ALF2,}  yield supramassive NSs in more than half of the mergers. However, for $j\lesssim 0.7j_{\rm Kep}$, the presence or absence of a non-negligible fraction of supramassive NSs formed in the mergers depends sensitively on both the EoS and the mass distribution of the binary systems. The NS mergers with a total gravitational mass $\leq  2.6M_\odot$ are found to be able to shed valuable light on both the EoS model and the angular momentum of the remnants if supramassive NSs are still absent. We have also discussed the uncertainty on estimating the maximum gravitational mass of non-rotating NSs ($M_{\rm max}$) due to the unknown $j$ of the pre-collapse remnants. With the data of GW170817 and the assumption of the mass loss of $0.03M_\odot$, we have $M_{\rm max}<(2.19,~2.32)M_\odot$ (90\% confidence level) for $j=(1.0,~0.8)j_{\rm Kep}$, respectively.
\end{abstract}

\keywords{gravitational wave---stars: neutron---binaries: close}

\section{Introduction} \label{sec:intro}
The long-standing hypothesis that the mergers of close NS binaries can generate energetic gravitational-wave (GW) \citep{1977ApJ...215..311C,LVC2010}, short gamma-ray bursts as well as a large amount of r-process material \citep{Eichler1989}, and Li-Paczy\'{n}ski macronova \citep{LiLX1998,2017LRR....20....3M} have been directly/convincingly confirmed by the successful detection of GW170817/GRB 170817A/AT2017gfo \citep{Abbott2017,von Kienlin2017,Pian2017,Covino2017}. The gravitational wave data of GW170817 also impose important constraints on the models of equation of state (EoS) of NSs and favor those yielding compact NSs, for instance APR4 and SLy \citep{Abbott2017}. Intriguingly, compact stars were also inferred from the
radius measurements based on the X-ray observations of a group of NSs \citep{Lattimer2016,Ozel2016}.

The EoS of NSs plays an important role in the fate of the merger remnants. In principle, the merger of a pair of NSs could  yield either a promptly-formed black hole, or a  hypermassive NS (HMNS) supported by the differential rotation, or a supramassive NS (SMNS) supported by its quick uniform rotation, or even a stable NS, depending sensitively on the EoS models and the total mass of the pre-merger binary stars \citep[e.g.][]{Morrison2004,Hotokezaka2013,Baiotti2017}. Before 2005, though the possibility of forming highly magnetized SMNS or stable NS (i.e., magnetars) was speculated in the literature \citep[e.g.][]{Duncan1992,Davis1994,Dai1998}, it is widely believed that the most likely outcome is either a black hole or a HMNS \citep[e.g.][]{Eichler1989,Duncan1992}, mainly because at that moment the accurate measurements of NSs only yield mass well below $\sim 2M_\odot$ (i.e., the EoS is likely soft) and there were no other observation data directly linked to NS mergers except the prompt emission of short GRBs.

The situation changed dramatically after the localization and hence the detection of the afterglow emission of short GRBs \citep[][]{Gehrels2005}. The ``unexpected" X-ray/optical flares or plateaus following the prompt emission of short GRBs  \citep{Fox2005,Villasenor2005} motivated some interesting ideas. Though the models of fallback accretion onto the nascent black hole have been discussed in the literature \citep{Fan2005,Perna2006,Rosswog2007,Lee2009}, more attention has been paid on the NS central engine possibility. In evaluating the possible optical emission associated with NS merger events, \citet{Kulkarni2005} proposed pioneeringly that a long-lived central engine can enhance the macronova emission significantly and a stable and relatively slowly-rotating magnetar was taken to yield the thermal optical/infrared emission of $\sim 10^{41}~{\rm erg~s^{-1}}$. To interpret the X-ray flares (also called the extended X-ray emission) following short GRB 050709 and GRB 050724, \citet{Dai2006} suggested that the central engines were long-lived differentially-rotating pulsars with a low dipole magnetic filed strength of $\sim 5\times 10^{10}$ Gauss and the outbursts were driven by the wrapping of the poloidal seed magnetic field to $\sim 10^{17}$ Gauss, while \citet{Gao2006} proposed that the central engines were supramassive magnetars formed in double NS mergers and the extended X-ray emission were due to the prompt magnetic energy dissipation of the Poynting-flux dominated winds. The energy injection from a supramassive magnetar was suggested by \citet{Fan2006} to account to the flat X-ray segment detected in short GRB 051221A  and these authors further proposed that the magnetar wind can accelerate the almost isotropic sub-relativistic ejecta launched during the NS merger to a mildly-relativistic velocity and then give rise to multi-wavelength afterglow emission. Intriguingly the numerical simulations do suggest that super-strong magnetic fields and hence magnetars could be generated/formed in double NS mergers \citep{Rosswog2006}. Subsequently the magnetar model has been adopted to interpret more sGRB data \citep[e.g.][]{Metzger2008,Rowlinson2010}. The magnatar  model has attracted much wider attention since 2013 when \citet{Rowlinson2013} claimed that about half of the X-ray afterglow emission of short GRBs are dominated by the radiation components from the magnetars. On the other hand, the accurate observations of pulsar PSR J0348+0432 have increased the lower limit of the maximum gravitational mass of a slowly-rotating NS ($M_{\rm max}$) to $2.01\pm0.04~M_\odot$ \citep{Antoniadis2013}, which significantly boosts the chance of forming long-lived remnants in NS mergers \citep[e.g.,][]{Morrison2004,Giacomazzo2013}.
 Then the ideas emerged in 2005-2006, including that the magnatar formed in double NS mergers can significantly enhance the macronova emission, or give rise to X-ray plateaus followed by abrupt drops and/or X-ray flares, or drive mildly-relativistic outflow and then yield long-lasting and largely isotropic afterglow emission, have been extensively investigated in the literature \citep[e.g.][]{Yu2013,Metzger2014,Zhang2013,Gao2013,Wu2014}.

The magnetar model for the X-ray plateau followed by an abrupt drop is however found to be challenged because the observed durations of
X-ray plateaus are significantly shorter than that expected in the magnetic dipole radiation scenario. The same problem arises in the magnetar energy injection model for the flat X-ray segment detected in GRB 130603B, a burst with a macronova signal in its late afterglow that has imposed additional tight constraints on the total energy output of the central engine. One possible solution for these two puzzles is that the superstrong toroidal fields have deformed the supramassive magnetars significantly and most of the kinetic rotational energy has been carried away by the gravitational wave in $\sim 10^{2}-10^{4}$ seconds \citep{Fan2013PRD,FYX2013,Gao2016,Lasky2016,Lv2017}. Moreover, the supramassive magnater model provides a simple way to directly estimate $M_{\rm max}$ \citep{Fan2013PRD,Lasky2014,Lv2014,LiX2014,Lawrence2015,Fryer,Gao2016,Margalit2017,Rezzolla2017}.

Though such progresses are encouraging, the fate of the remnants formed in NS stars is still uncertain due to the lack of a well determined EoS model of NSs \citep{Oertel2017} and the poorly-constrained energy loss of the newly-formed supermassive NSs in their differential rotation phase. Nevertheless, one can carry out the EoS model-dependent NS merger numerical simulations and then estimate the fate of the remnants. For example, \citet{Morrison2004} investigated the mergers of the Galactic systems and concluded that for APR4 like models, SMNSs are the likely outcome \citep[see also][]{Piro2017}. The actual fate however likely depends on the non-equilibrium dynamics of the coalescence event and the rapid kinetic rotational energy loss \citep{Lawrence2015}. { Due to the lack of widely-accepted constraints, these previous works usually adopted a few representative models to cover the soft, middle and stiff EoSs. For example, \citet{Piro2017} took the models of H4, APR4, GM1, MS1 and SHT, while \citet{Fryer} adopted the EoS models of NL3, FSU2.1 and one that fits some neutron star observation data. As mentioned in the beginning, the gravitational wave data of GW170817 favor the EoSs yielding compact neutron stars \citep[see Fig.5 of][]{Abbott2017} and the spectroscopic radius measurements of a group of NSs draw the same conclusion \citep{Ozel2016}. Motivated by such intriguing progresses, in this work we re-examine the fate of the NS merger remnants and focus on the EoSs that are not stiffer than H4. In some literature the merger remnants are assumed to be near the mass-shedding limit (i.e., $j\approx j_{\rm Kep}$). We instead assume that efficient angular momentum loss in the merger is possible and $j\leq j_{\rm Kep}$. The roles of the angular momentum and the mass distribution of the binary NSs on shaping the remnants' fate are investigated. This work is structured as the following: in Section 2 we introduce a simplified approach to estimate whether the formed remnant is a supramassive NS or not. In Section 3 we
evaluate the fate of the remnants formed in double NS mergers with some EoS models predicting compact stars. In Section 4 we discuss the possible constraints on $j/j_{\rm Kep}$ with GW170817 and the uncertainty induced on $M_{\rm max}$. Finally we summarize our results with some discussions.
}

\section{The condition for forming SMNS in NS mergers: the simplified approach}\label{sec:formulae}
\citet{Beru2016}  found that the binding
energy (BE) of a NS, defined as the difference between the baryonic mass ($M_{\rm b}$) and the gravitational mass ($M$) of an equilibrium configuration,
could show a behaviour that is essentially
independent of the EOS, which reads \citep[see also][]{Lattimer2001}
\begin{equation}
{{\rm BE}\over M}=0.619\zeta+0.136\zeta^{2},
\label{eq:BE}
\end{equation}
where the compactness $\zeta=GM/Rc^{2}$ and $R$ is the radius of the NS. Therefore, for a given EOS the BE can be directly calculated.

The pre-collapse remnant formed in the merger of a pair of NSs has a baryonic mass of
\begin{equation}
M_{\rm b}=M_{1}+M_{2}+{\rm BE}_{1}+{\rm BE}_{2}-m_{\rm loss},
\label{eq:total-baryonic}
\end{equation}
where $M_1$ and $M_2$ are the gravitational masses of the pre-merger NSs (their summary is the so-called total gravitational mass, i.e., $M_{\rm tot}=M_1+M_2$) and $m_{\rm loss}$ is the rest mass of the material lost in the merger process, including the dynamical ejecta and the neutrino-driven winds from the accretion disk and the pre-collapse remnant.

On the other hand, for a given EoS, the non-rotating NS can only support the baryonic mass of
\begin{equation}
M_{\rm b,max}=M_{\rm max}+{\rm BE}_{\rm max}.
\end{equation}
For $M_{\rm b}\leq M_{\rm b,max}$ we have stable NSs.

To form  SMNSs that are supported by quick uniform rotation, it is required that
\begin{equation}
M_{\rm tot}+{\rm BE}_{1}+{\rm BE}_{2}-m_{\rm loss}\leq M_{\rm b,crit},
\label{eq:Constraint}
\end{equation}
where $M_{\rm b,crit}=M_{\rm crit}+{\rm BE}_{\rm crit}$, for $M_{\rm crit}$
we adopt the simple polynomial fitting function of the enhancement of the rigid rotation on the maximum gravitational mass, which reads \citep[see also][for similar approaches]{Friedman1986,Lasota1996}
\begin{equation}
{M_{\rm crit}\over M_{\rm max}}=1+0.1316(j/j_{\rm Kep})^{2}+0.0711(j/j_{\rm Kep})^{4},
\label{eq:Mcrit}
\end{equation}
where $j$ is the dimensionless angular
momentum and $j_{\rm Kep}$ is the maximum value
allowed for a given EOS \citep{Beru2016}. For $j=j_{\rm Kep}$ (i.e., the mass shedding limit), we have $M_{\rm crit}\approx 1.2M_{\rm max}$. While for
$j=(0.8,~0.6)j_{\rm Kep}$, we will have $M_{\rm crit}\approx (1.1,~1.06)M_{\rm max}$. In view of the narrow distribution of the masses of the binary NSs found in the Galaxy, such a large difference may have a very important effect on the fate of the merger remnants, which is the focus of this work.

Following \citet{Beru2016} we approximate the binding
energy of a quickly-rotating star with a gravitational mass of $M_{\rm crit}$ as
\begin{equation}
{{\rm BE}_{\rm crit}\over M_{\rm crit}}=0.619(1-1.966\times10^{-2}j+4.272 \times 10^{-1}j^{2})\zeta+0.136(1+4.46\times10^{-1}j-7.603j^{2})\zeta^{2}.
\label{eq:SMNS-1}
\end{equation}
The above equation can be re-expressed into the term of $j/j_{\rm Kep}$ (i.e., to be in the same form of eq.(\ref{eq:Mcrit})) by taking into account the approximation of \citet{Beru2016}
\begin{equation}
j=j_{\rm Kep}(j/j_{\rm Kep})\approx 0.5543(1/\zeta)^{1/2}(j/j_{\rm Kep}).
\label{eq:SMNS-2}
\end{equation}
Note that in eq.(\ref{eq:SMNS-1}) and eq.(\ref{eq:SMNS-2}) the term of $\zeta$ is that of the non-rotating NS with a gravitational mass of $M_{\rm max}$.

Therefore, for the given EoS, $M_{1}$ and $M_2$, the fate of the formed remnant mainly depends on the value of $j/j_{\rm Kep}$. If the kinetic rotational energy of the newly formed remnants carried away by the gravitational wave radiation and the neutrinos (note that in the process of breaking the differential-rotation of the remnant,  some kinetic rotational energy will be converted into thermal energy and then carried away by the neutrinos) is relatively inefficient in the differential rotation phase, one would expect that $j/j_{\rm Kep}\approx 1$ \citep[][i.e., the mass-shedding limit]{Lawrence2015,Piro2017}. If instead the relevant kinetic energy losses are efficient \citep[e.g.,][]{Sekiguchi2011,Hotokezaka2013,Bernuzzi2016,ZhangXF2017}, $j$ may be sizably below $j_{\rm Kep}$. In view of such uncertainties, in this work we assume a wide range of $j/j_{\rm Kep}=(0.6,0.7,0.8,0.9,1.0)$.

\section{The fate of the remnants formed in double NS mergers: some EoS models yielding compact stars}
The gravitational wave data of GW170817
favor the equations of states that
predict compact NSs \citep{Abbott2017}, while the constraints from X-ray
observations of a group of NSs yield the same conclusion \citep{Lattimer2016,Ozel2016,Bogdanov2016}. Motivated by such intriguing progresses and the fact that $M_{\rm max}\geq 2.01\pm0.04~M_\odot$ \citep{Antoniadis2013}, in this work we consider seven types of ``currently-favored" EoS models, including
APR4 \citep{Akmal1998}, SLy \citep{Douchin2001}, {ALF2 \citep{Alford2005}, H4 \citep{Glendenning1991}, ENG \citep{Engvik1996}, MPA1 \citep{MPA1987}} and one specific empirical model based on current NS radius measurements \citep[herefater the ``NS radius based EoS", the $M-R$ relation needed in our approach is adopted as the solid blue curve in the left panel of Fig.10 of][]{Bogdanov2016}. The main purpose of this work is to evaluate the prospect of forming SMNSs in the double NS mergers. As mentioned above, one of the main uncertainties is the rotational kinetic energy of the newly formed remnants carried away by the gravitational wave radiation and the neutrinos in the differential rotation phase. In the following approach we assume $j/j_{\rm Kep}=0.6,0.7,0.8,0.9,1.0$, respectively.
If eq.(\ref{eq:Constraint}) is satisfied, SMNSs are formed, otherwise black holes are the outcome.

So far, the observed binary NS systems are still handful \citep{Lattimer2012}. Even for such a very limited sample, the masses of individual NSs in some binaries are not accurately measured. Therefore, the mass distribution of the NSs is still to be better determined. With a rather high NS merger rate \citep{Abbott2017,Jin2017}, such distributions may be mainly measured with the gravitational wave data in the next decade (please note that in the advanced LIGO/Virgo era, the $M_{\rm tot}$ of the per-merger binaries rather than the mass of the individual stars can be accurately measured). For the current purpose we adopt { three} representative distribution models based on the fit of the Galactic NS data.
The first is the double Gaussian distribution model. For the binary NS systems, \citet{Ozel2012} divided the sample into one of pulsars and
one of the companions (For the double pulsar system J0737-3039A, they
assigned the faster pulsar to the ``pulsar" and the slower to
the ``companion" categories). Repeating the above inference
for these two subgroups individually, \citet{Ozel2012} took the NS gravitational mass distribution as
$dN_{\rm NS}/dM \propto \exp[-(M-M_0)^2/2\sigma^2]$ and found that the most
likely parameters of the mass distribution for the pulsars are
$M_0 = 1.35M_\odot$ and $\sigma = 0.05~M_\odot$, whereas for the companions
$M_0 = 1.32M_\odot$ and $\sigma= 0.05M_\odot$. The second is the single Gaussian distribution model drawn from the galactic populations, which has a mean mass of $M_0=1.32 M_\odot$ and a standard deviation of $\sigma=0.11 M_\odot$ \citep{Kiziltan2013}.  Clearly, the second model has a wider mass distribution. { Different from these two models based on the Galactic NS observations, the third is from the population synthesis calculations by \citet{Fryer}. These authors calculated the mass distribution of the double neutron star systems for two metallicities: $Z=0.02$ (i.e., the high metallicity model) and $Z=0.002$ (i.e., the low metallicity model) and used the evenly mixed population of the high and low metellicity models  to approximately mimic a stellar content at various redshifts. In this work, such an approach is also adopted.}

With a given EoS and a NS mass distribution model, it is straightforward to calculate the baryonic mass of the remnant with eq.(\ref{eq:total-baryonic}) and hence estimate the fate of the remnant with eqs.(\ref{eq:Constraint}-\ref{eq:SMNS-2}), where $m_{\rm loss}$ is evaluated below.
{ As found by \citet{Hotokezaka2013} and \citet{Dietrich2017}, for the EoSs as soft as APR4 and SLy, $m_{\rm loss}$ depends insensitively on the mass ratio
of the two NSs (i.e., $q$); while for the stiffer EoSs, a sensitive dependence of $m_{\rm loss}$ on $q$ does present \citep[see also][]{Bauswein2013}. Recently, \citet{Dietrich2017b}
 used a large set of numerical relativity data obtained from different groups to derive phenomenological
fits relating the binary parameters to the ejecta properties. Intriguingly, these authors found an empirical relation between the dynamical ejecta mass and the physical
properties of the double neutron stars (see eq.(1) therein). Such a relation is directly adopted in our following approach for the EoSs yielding a radius $R_{1.35M_\odot}\geq 12$ km, where $R_{1.35M_\odot}$ represents the radius of the neutron star that is 1.35 times more massive than the sun. For the softer EoSs (i.e., APR4 and SLy), such a relation needs to be adjusted to better match the data.
Eq.(1) of \citet{Dietrich2017b} consists of two parts, including a constant term and a function of the neutron star properties.
To better reproduce the simulation data of APR4 and SLy models reported in \citet{Hotokezaka2013} and \citet{Dietrich2017} respectively, we further multiply these two parts by different coefficients (i.e., ${\cal K}_1$ and ${\cal K}_2$). For APR4 and SLy we take $({\cal K}_{1},~{\cal K}_2)=(1.17,~0.29)$ and $(0.61,~0.15)$, respectively.
In addition to the dynamical mass ejection, the neutrino-driven wind launches non-ignorable outflow \citep{2017LRR....20....3M}. Therefore, a further mass loss $\sim 0.01M_\odot$ is added to the dynamical ejecta to get the final $m_{\rm loss}$. For the NS radius-based EoS that is softer than both APR4 and SLy, so far no numerical simulation has been carried out. Since the soft EoSs are found to be insensitively dependent of $q$, we simply take $m_{\rm loss}\sim 0.03M_\odot$ and $0.05M_\odot$ (i.e., to match the mass range found in the modeling of AT2017gfo by \citet{Pian2017}), respectively, in our evaluation.
}

\begin{figure}[h]
\centering
\includegraphics[width=0.3\textwidth]{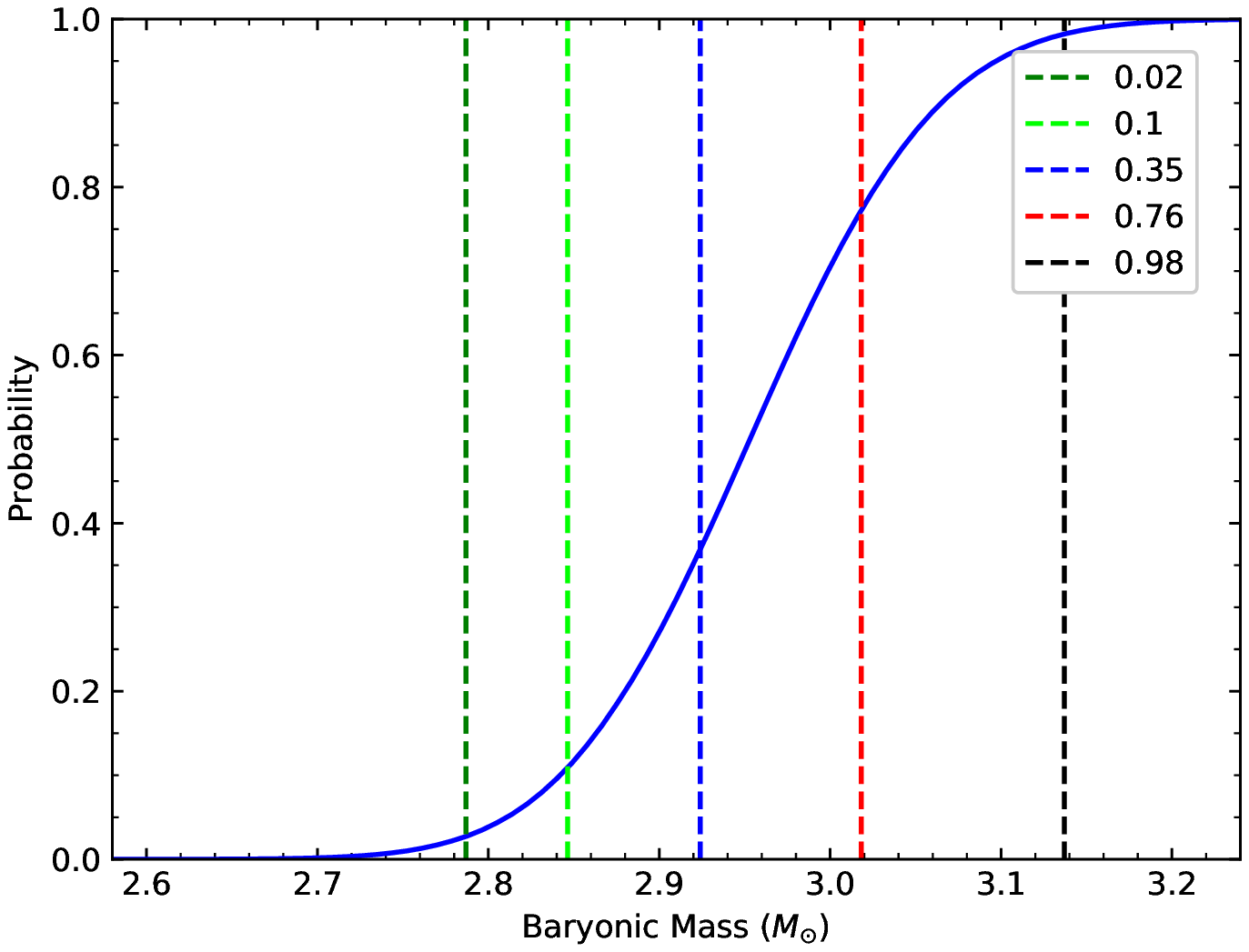}
\includegraphics[width=0.3\textwidth]{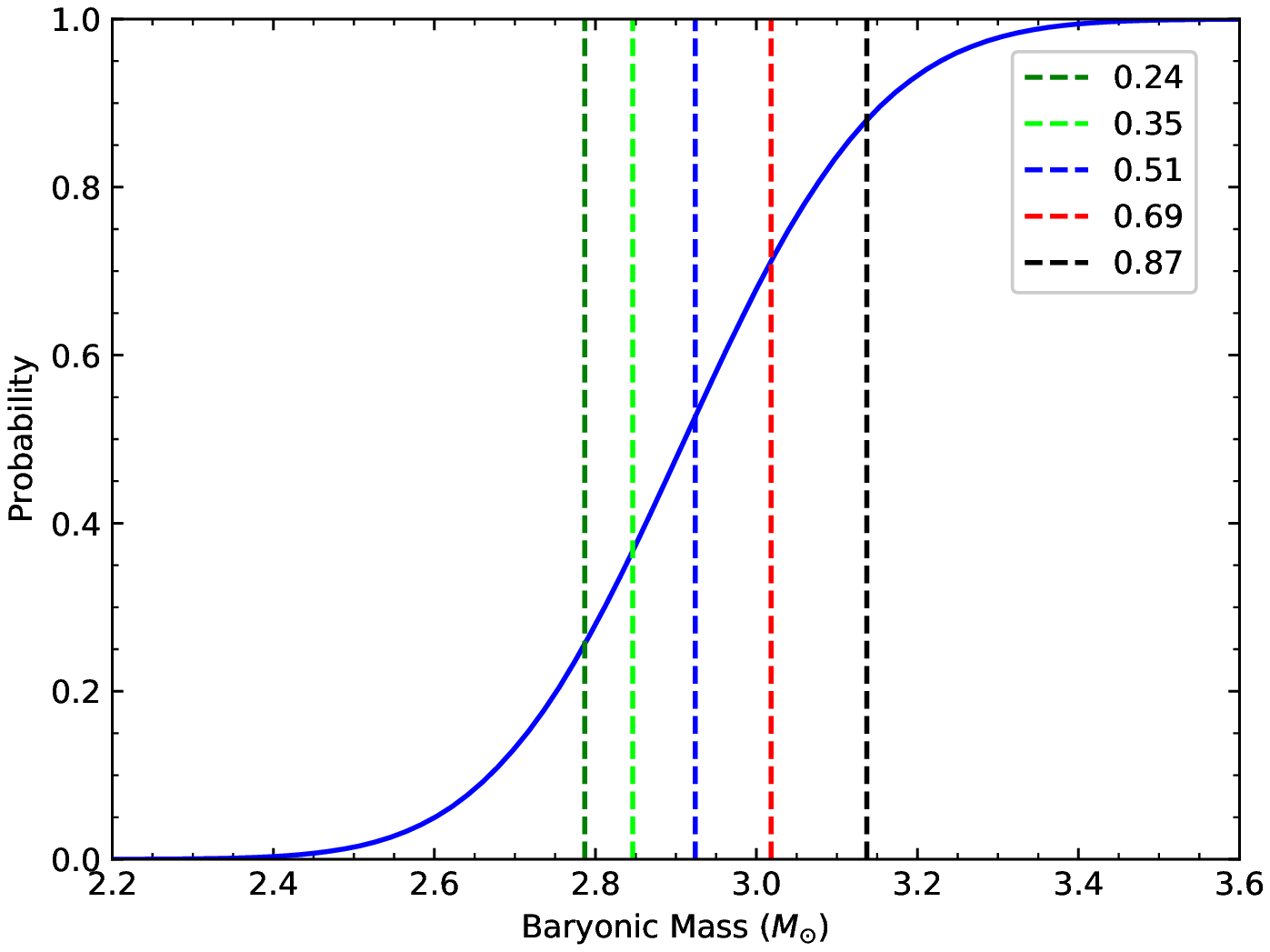}
\includegraphics[width=0.3\textwidth]{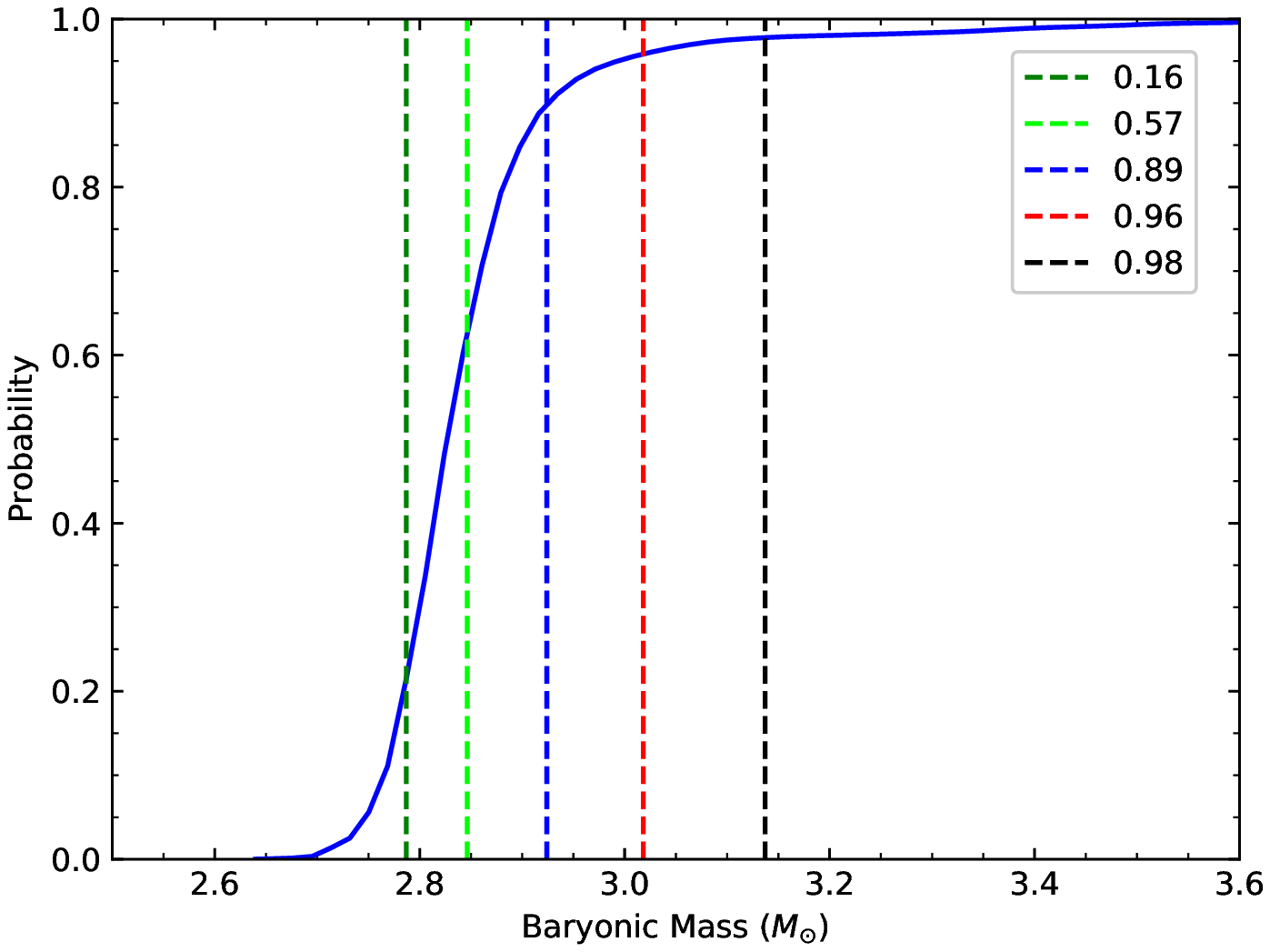}
\includegraphics[width=0.3\textwidth]{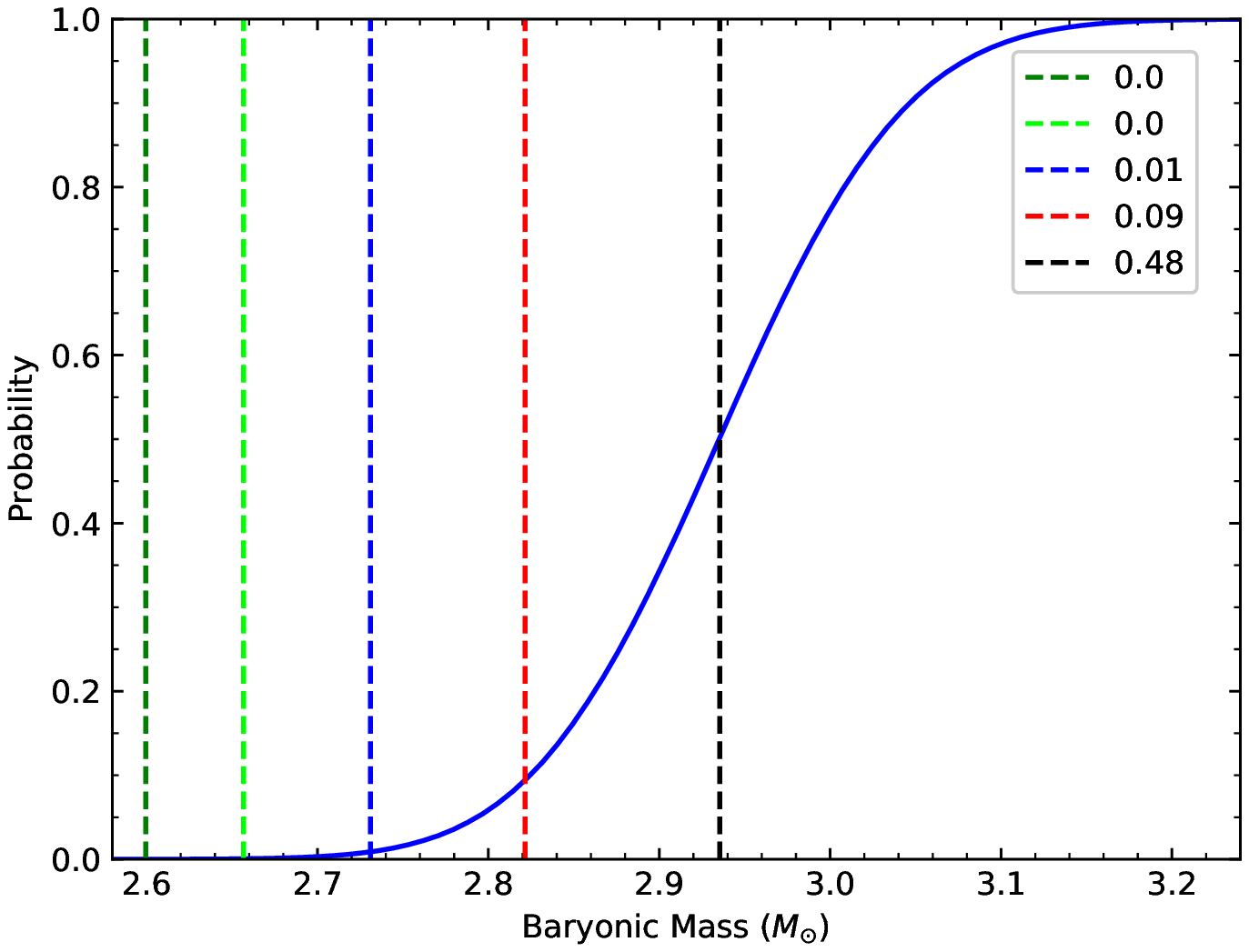}
\includegraphics[width=0.3\textwidth]{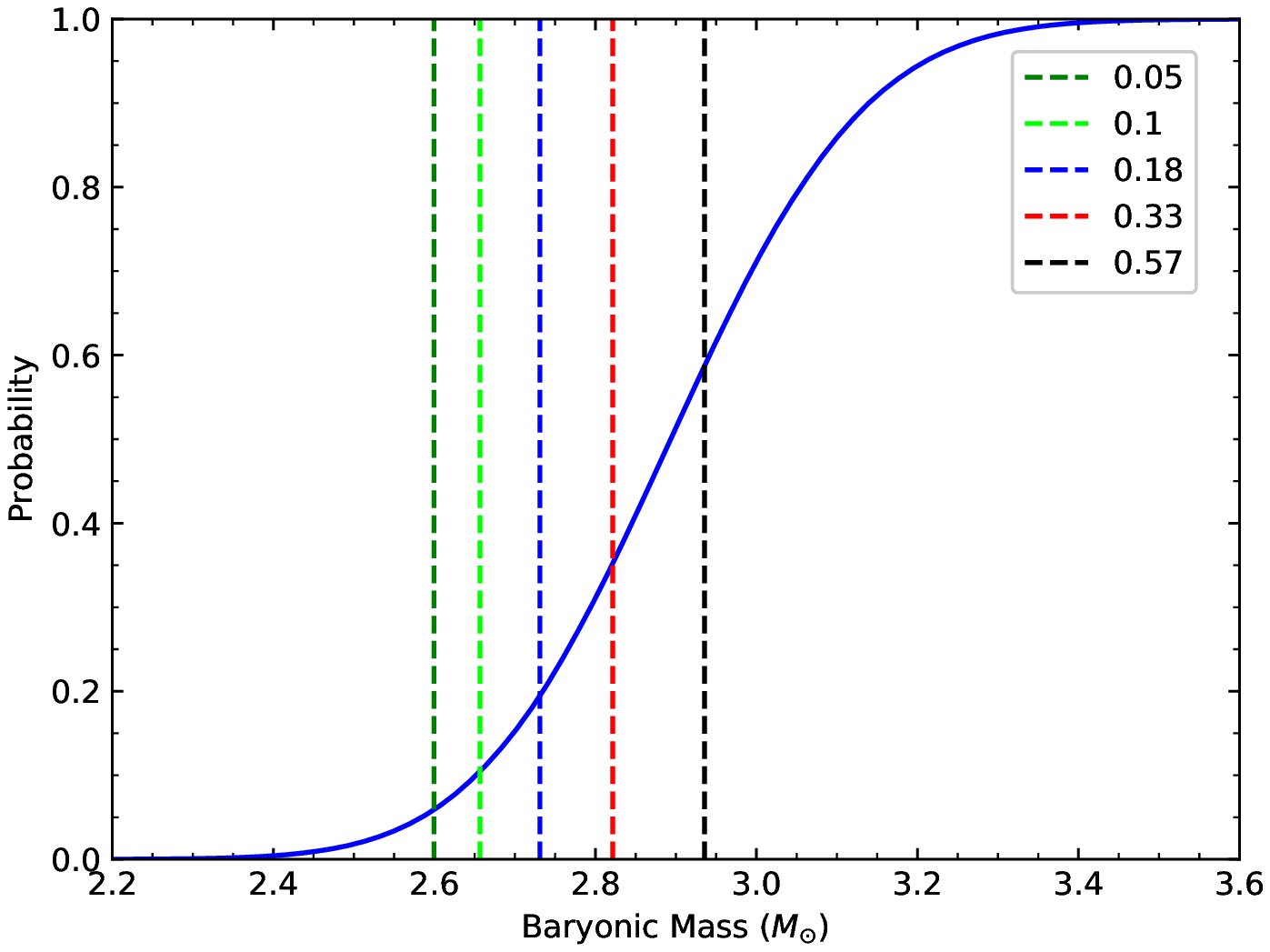}
\includegraphics[width=0.3\textwidth]{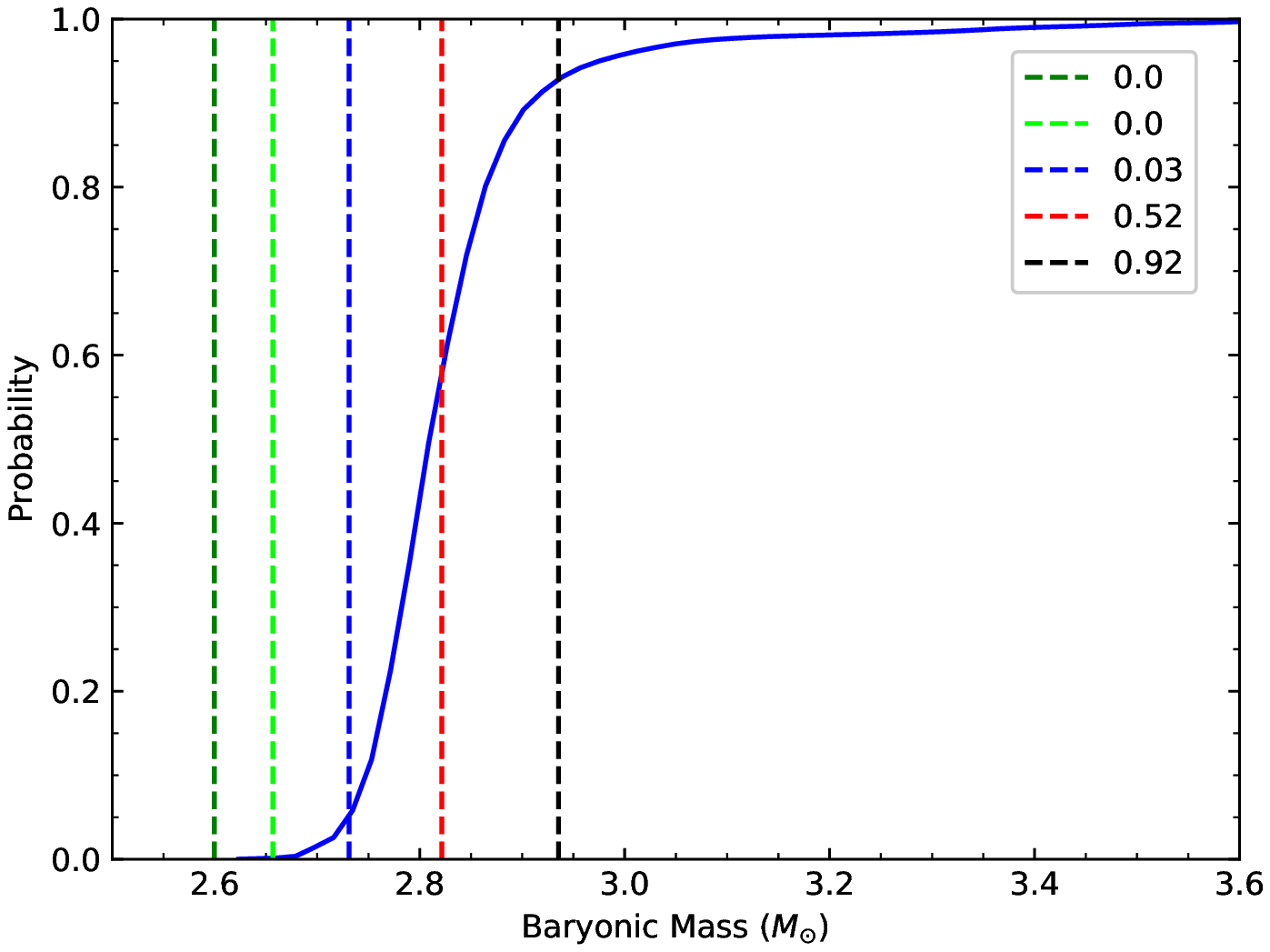}
\includegraphics[width=0.3\textwidth]{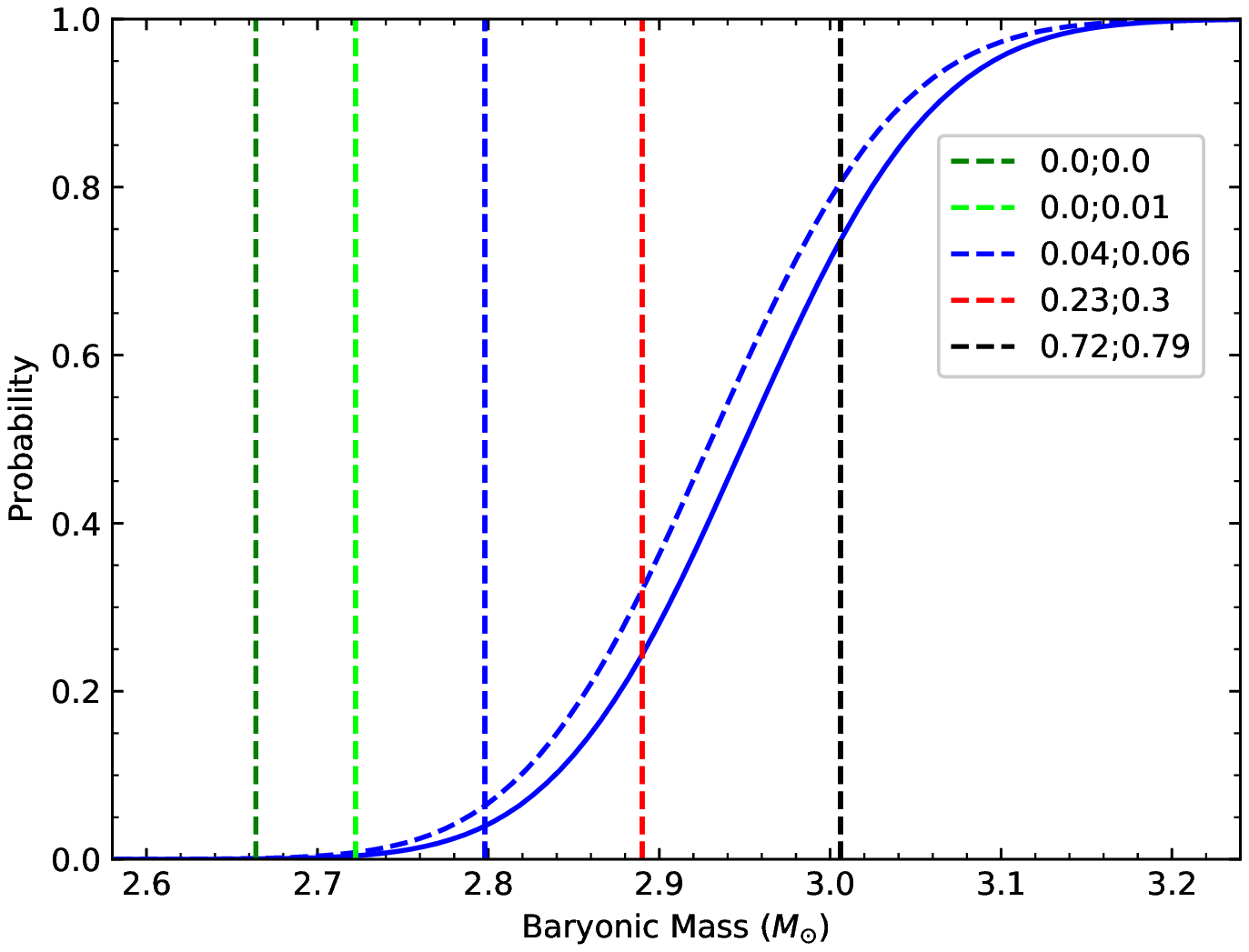}
\includegraphics[width=0.3\textwidth]{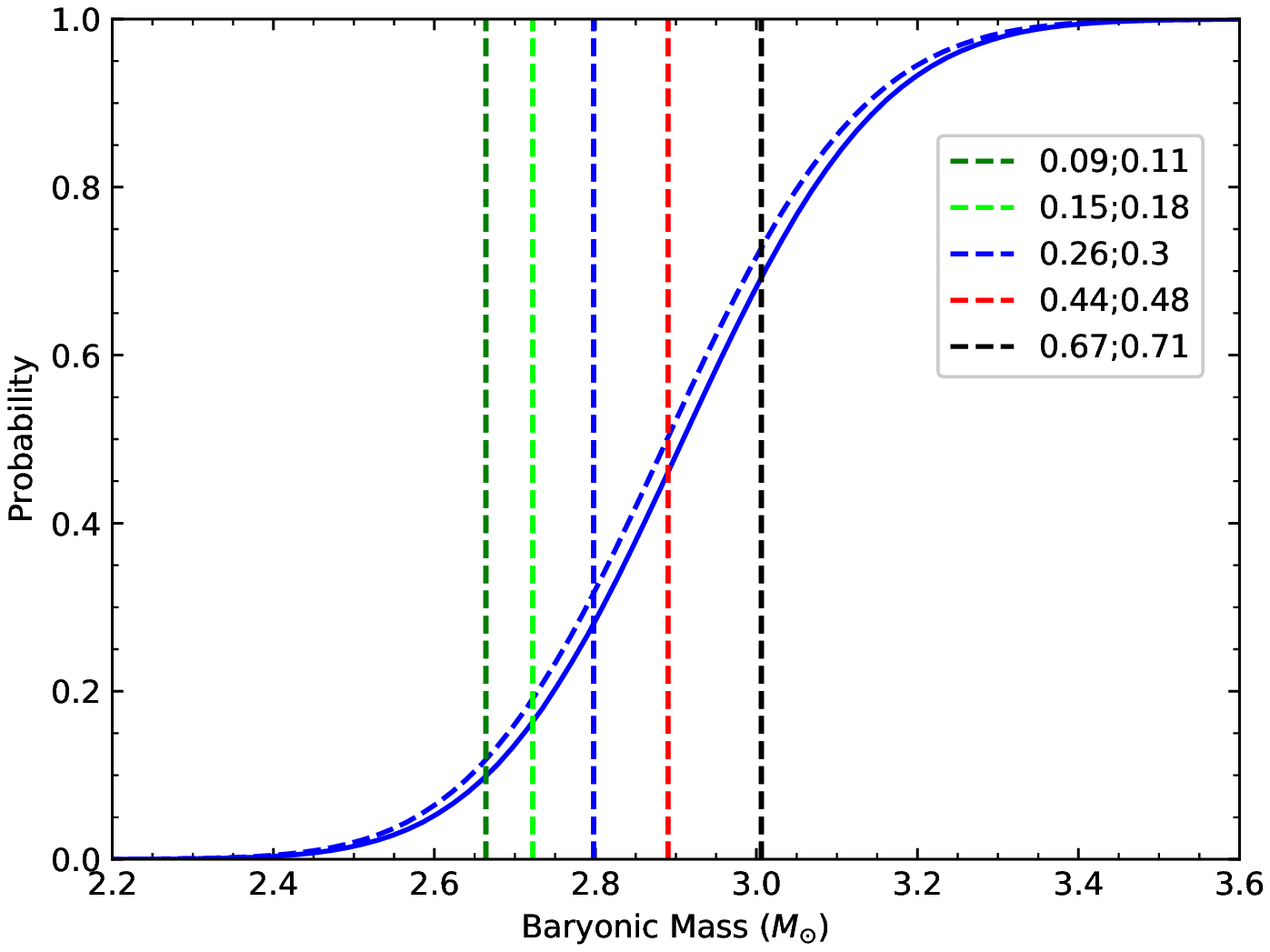}
\includegraphics[width=0.3\textwidth]{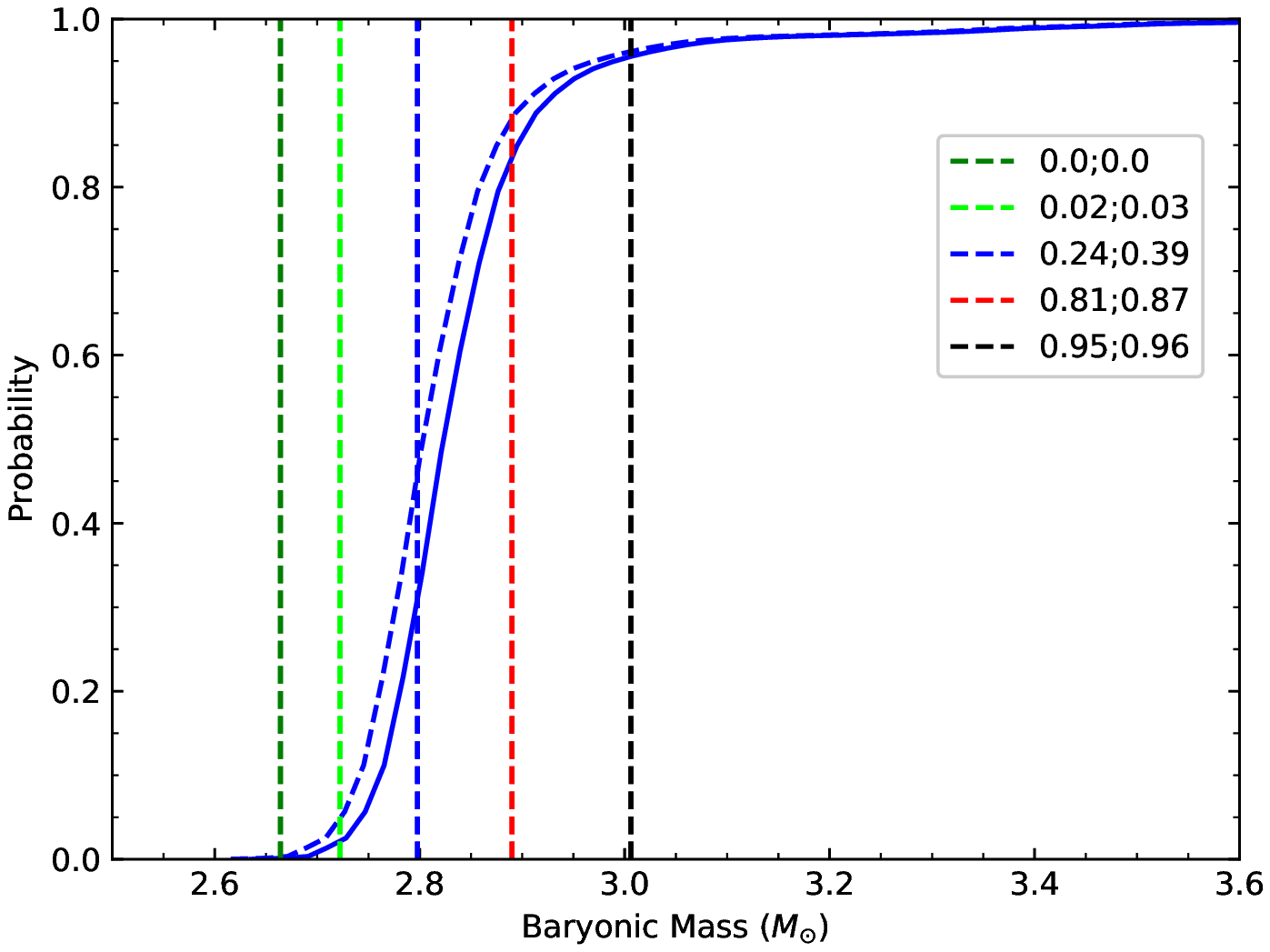}
\caption{{The expected distribution possibility of the baryonic mass of the remnants formed in double NS mergers. The top, middle and bottom panels are for APR4, SLy and the NS radius-based EoS models, respectively. For the (left, middle, right) column, the (double Gaussian, single Gaussian, population synthesis based) mass distribution model has been adopted from \citet{Ozel2012}, \citet{Kiziltan2013} and \citet{Fryer}, respectively. The vertical lines, from left to right, represent $j/j_{\rm Kep}=(0.6,~0.7,~0.8,~0.9,~1.0)$, respectively.
{ The two curves in the bottom panels are for $m_{loss} =0.03M_{\odot}$ (the dashed line) and $0.05M_{\odot}$ (the solid line), respectively.}
The possibility of forming SMNSs (sometimes including a small amount of stable NSs) in double NS mergers in each case is marked in the legend.
}} \label{fig:simulation-distribution}
\end{figure}

\begin{figure}
\centering
\includegraphics[width=0.3\textwidth]{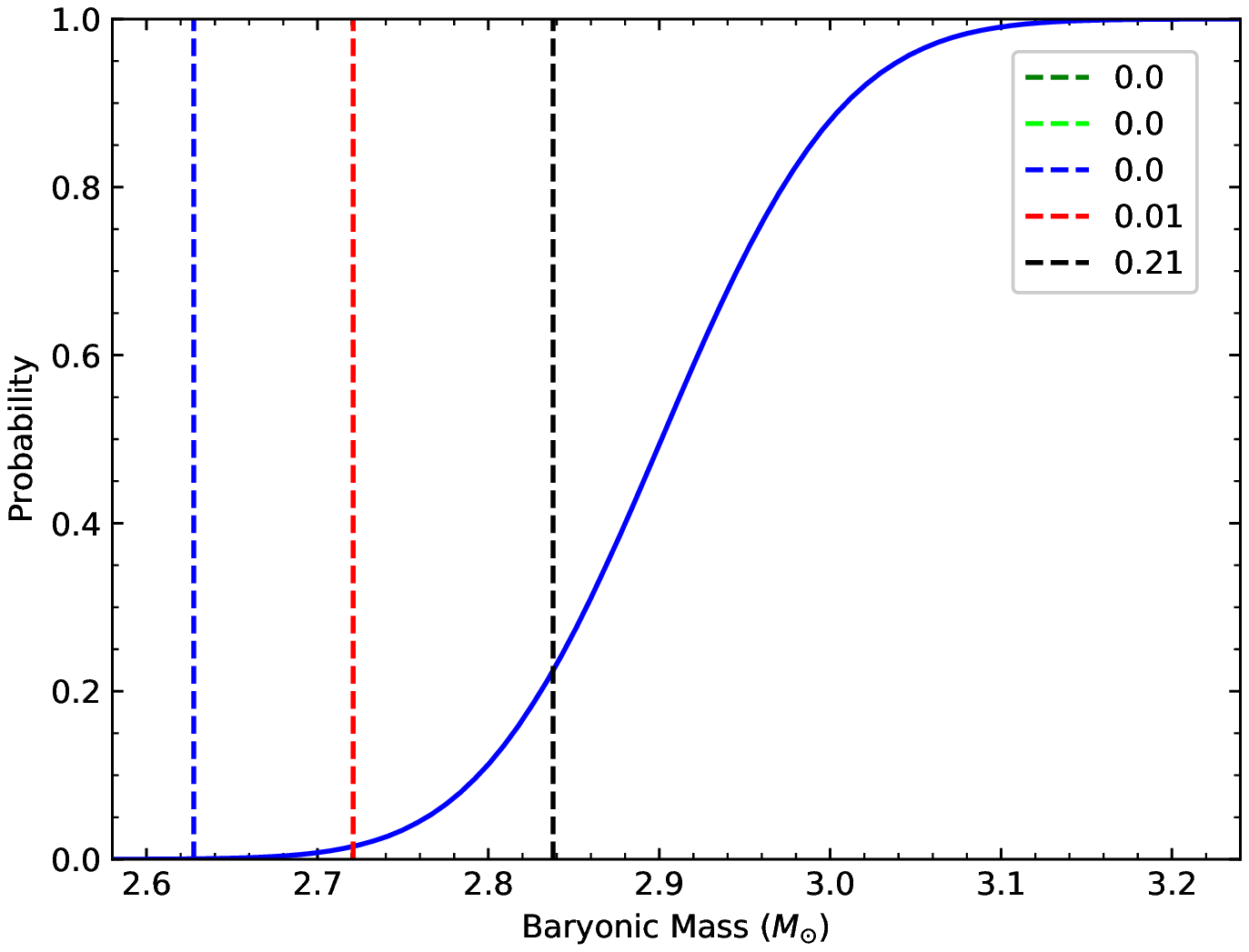}
\includegraphics[width=0.3\textwidth]{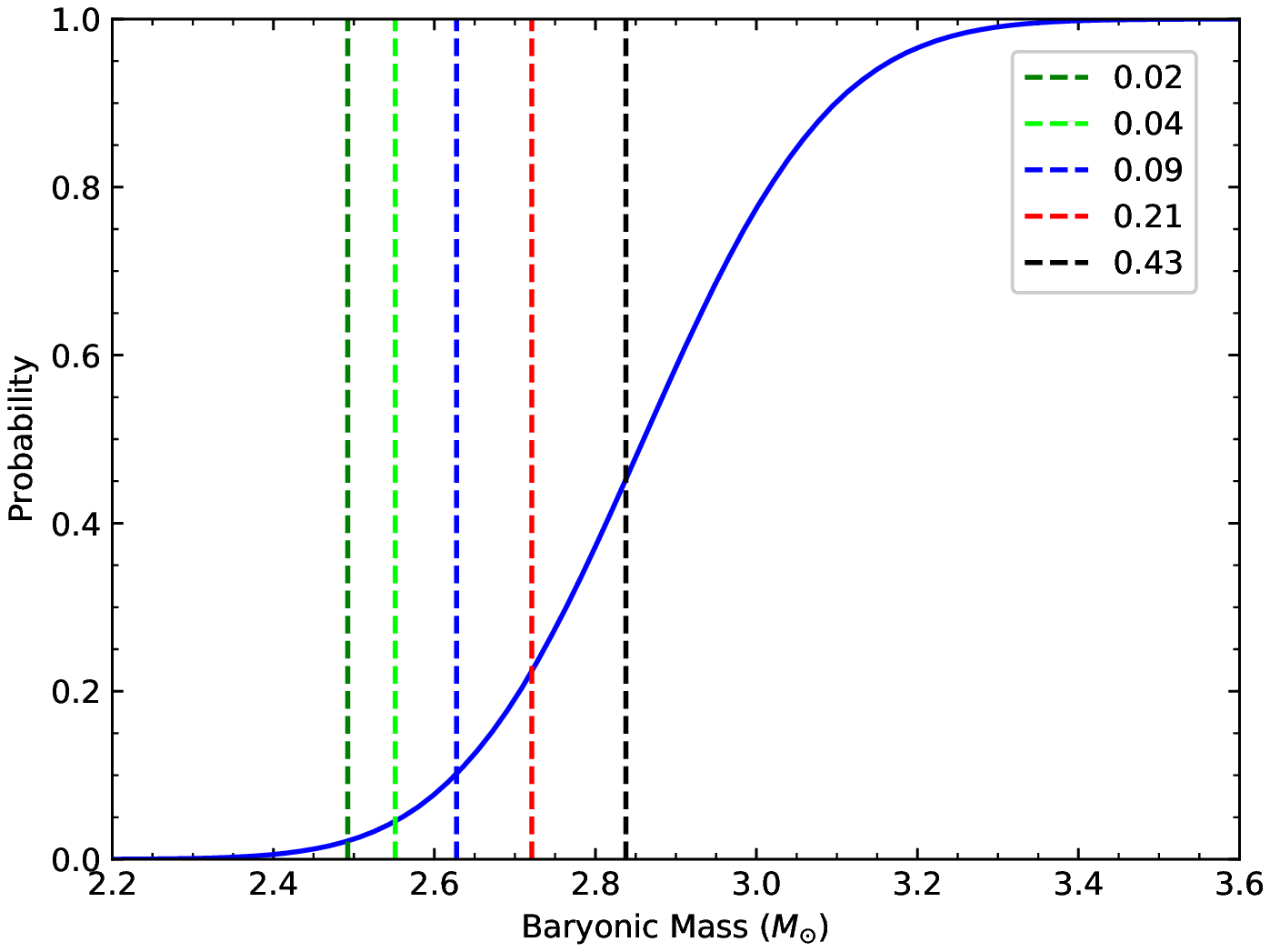}
\includegraphics[width=0.3\textwidth]{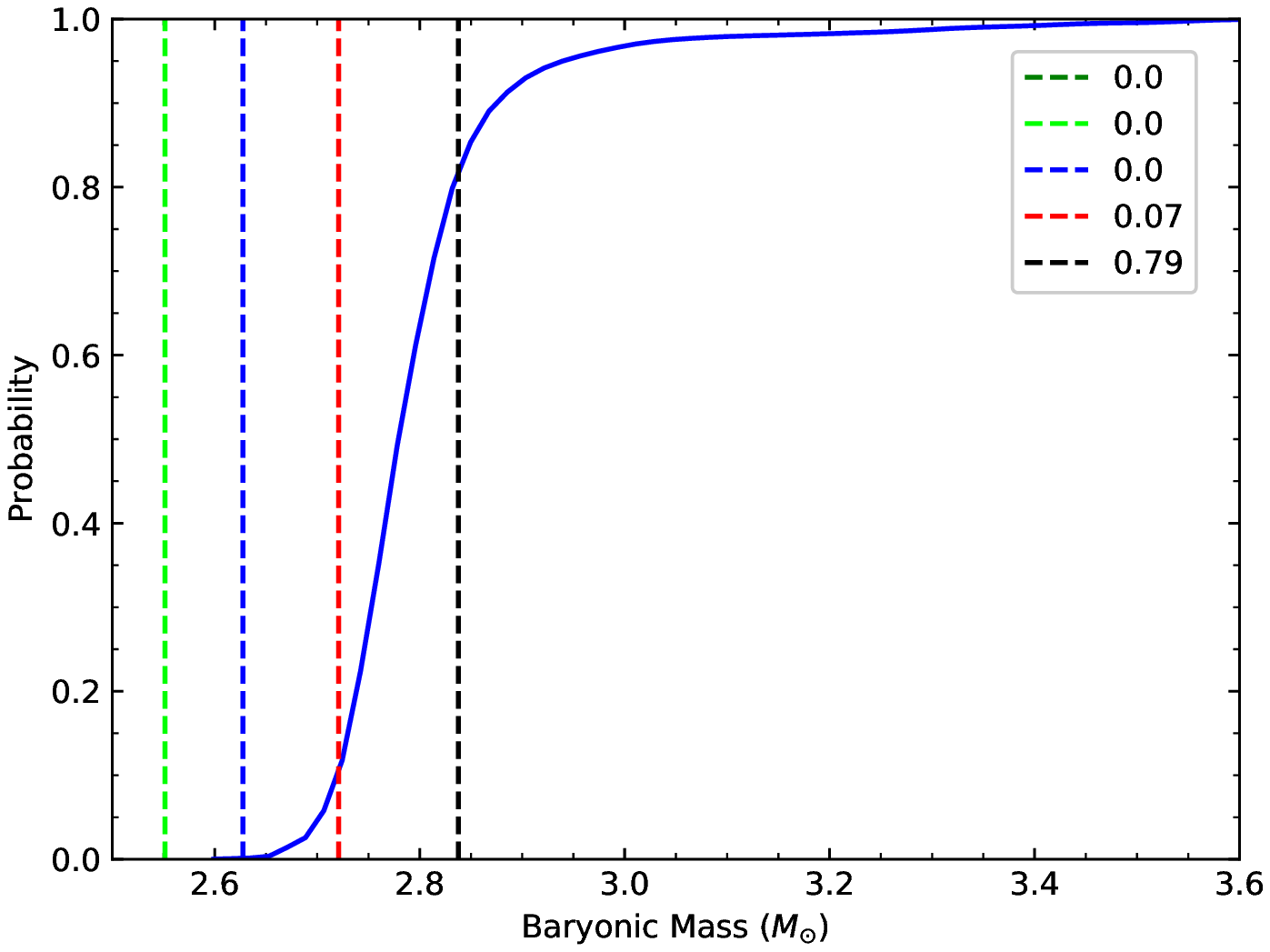}
\includegraphics[width=0.3\textwidth]{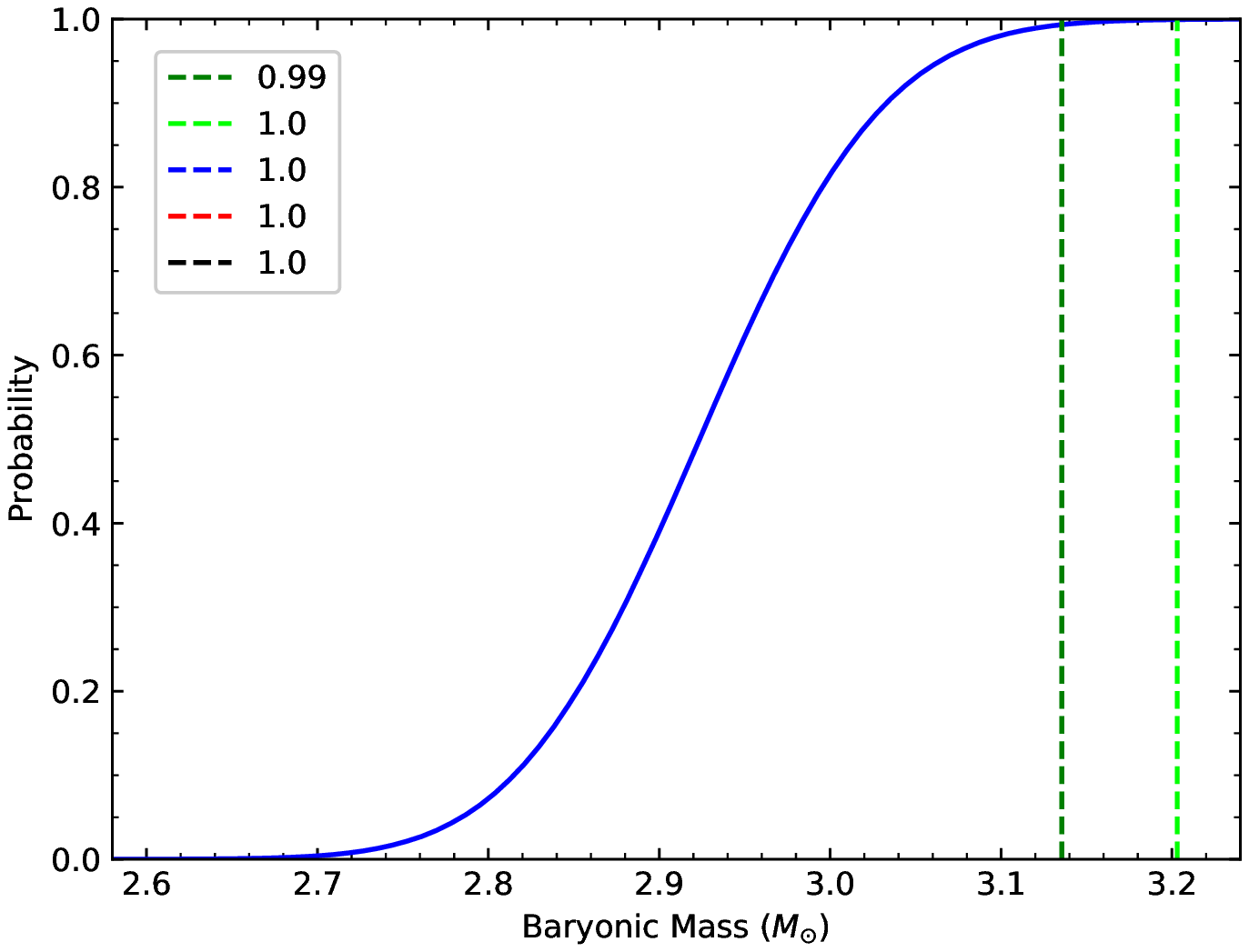}
\includegraphics[width=0.3\textwidth]{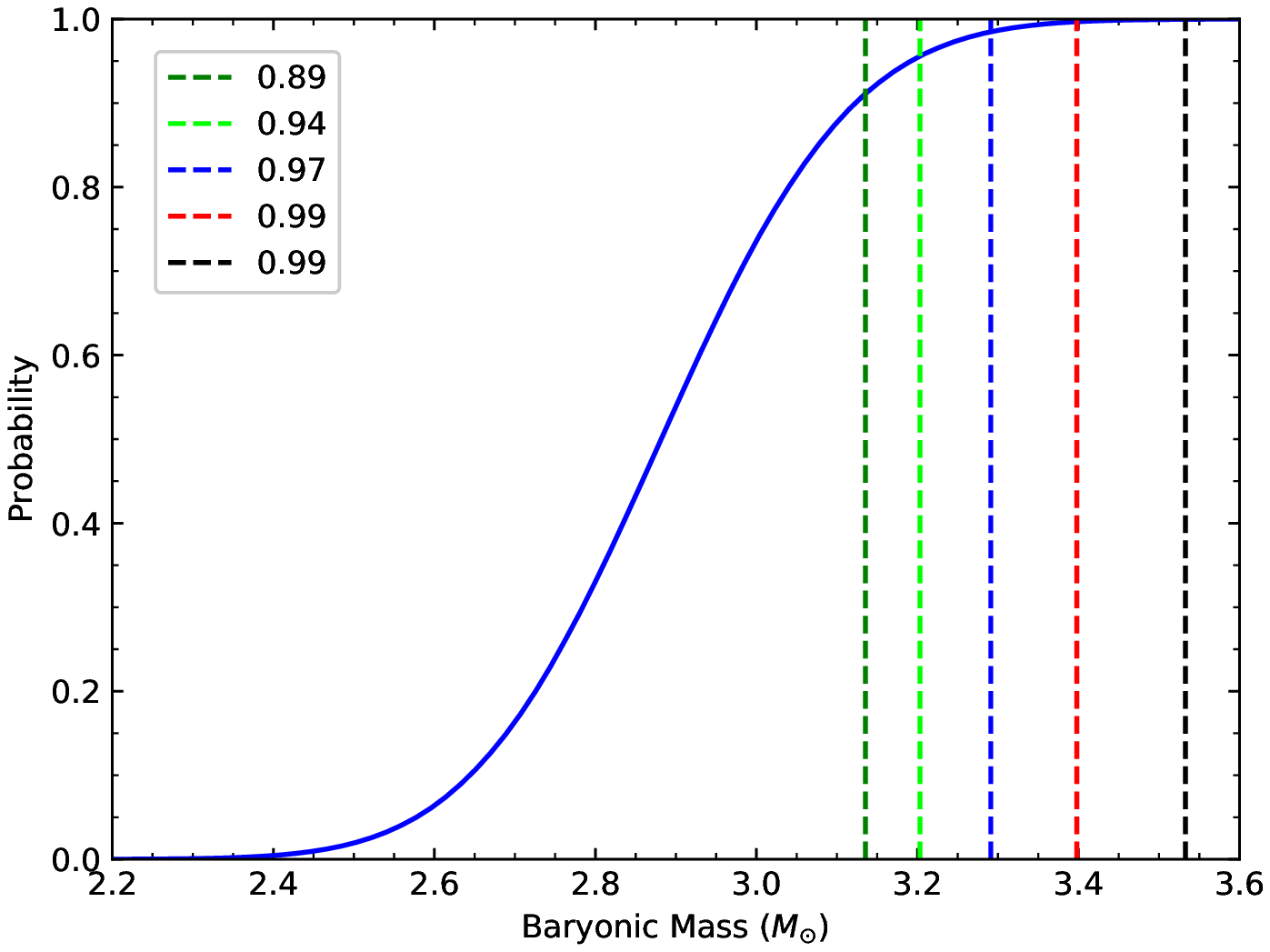}
\includegraphics[width=0.3\textwidth]{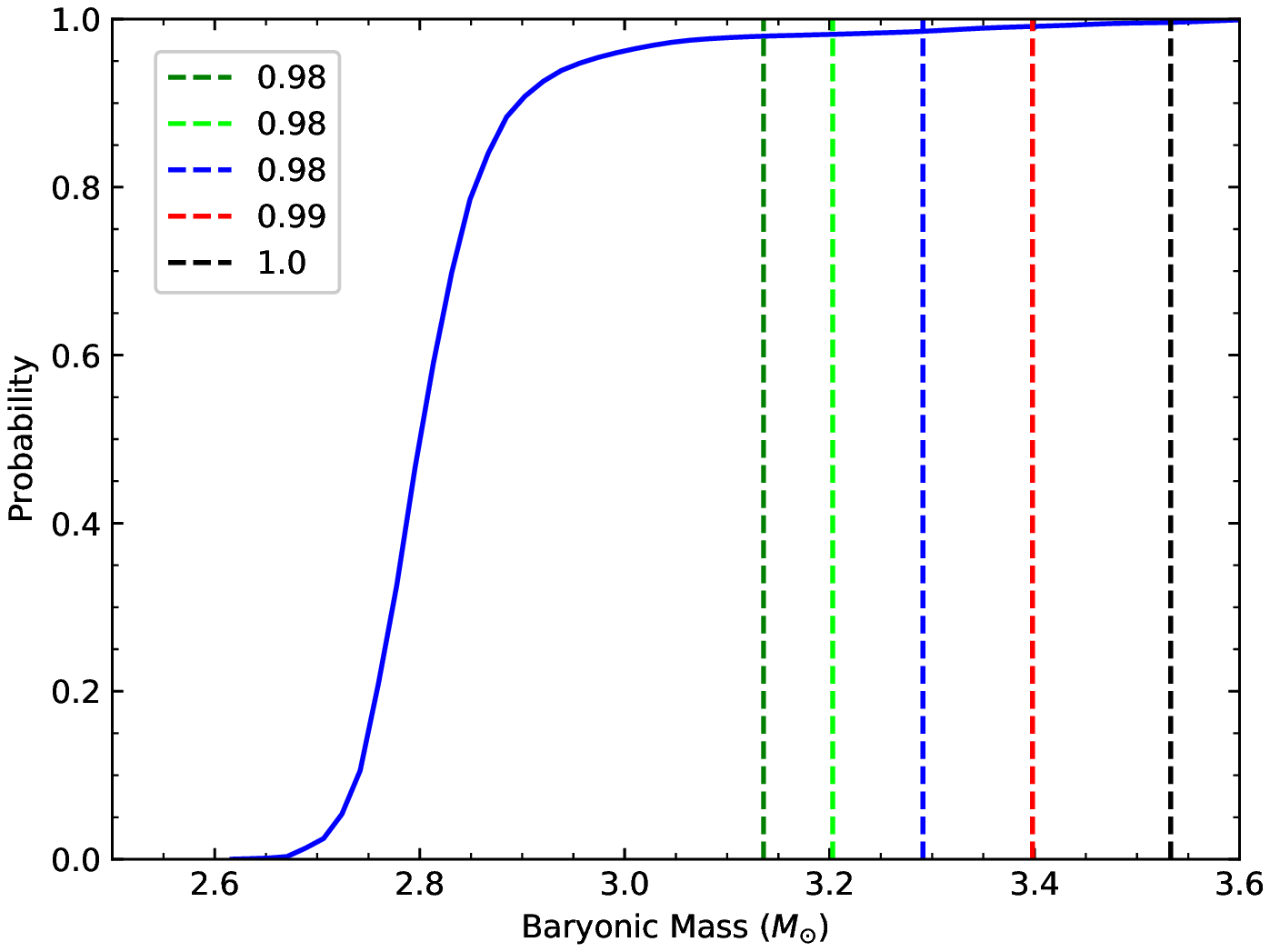}
\includegraphics[width=0.3\textwidth]{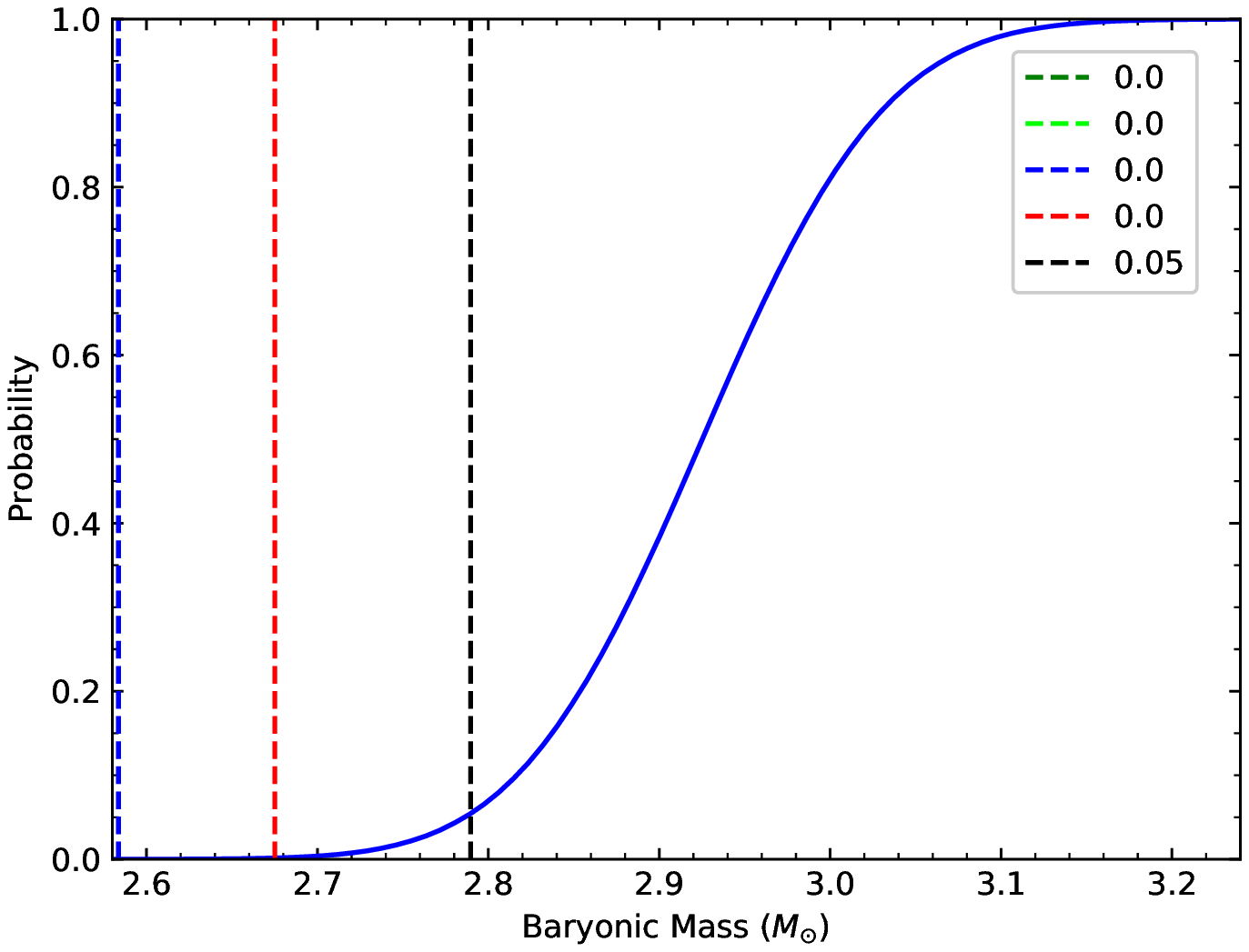}
\includegraphics[width=0.3\textwidth]{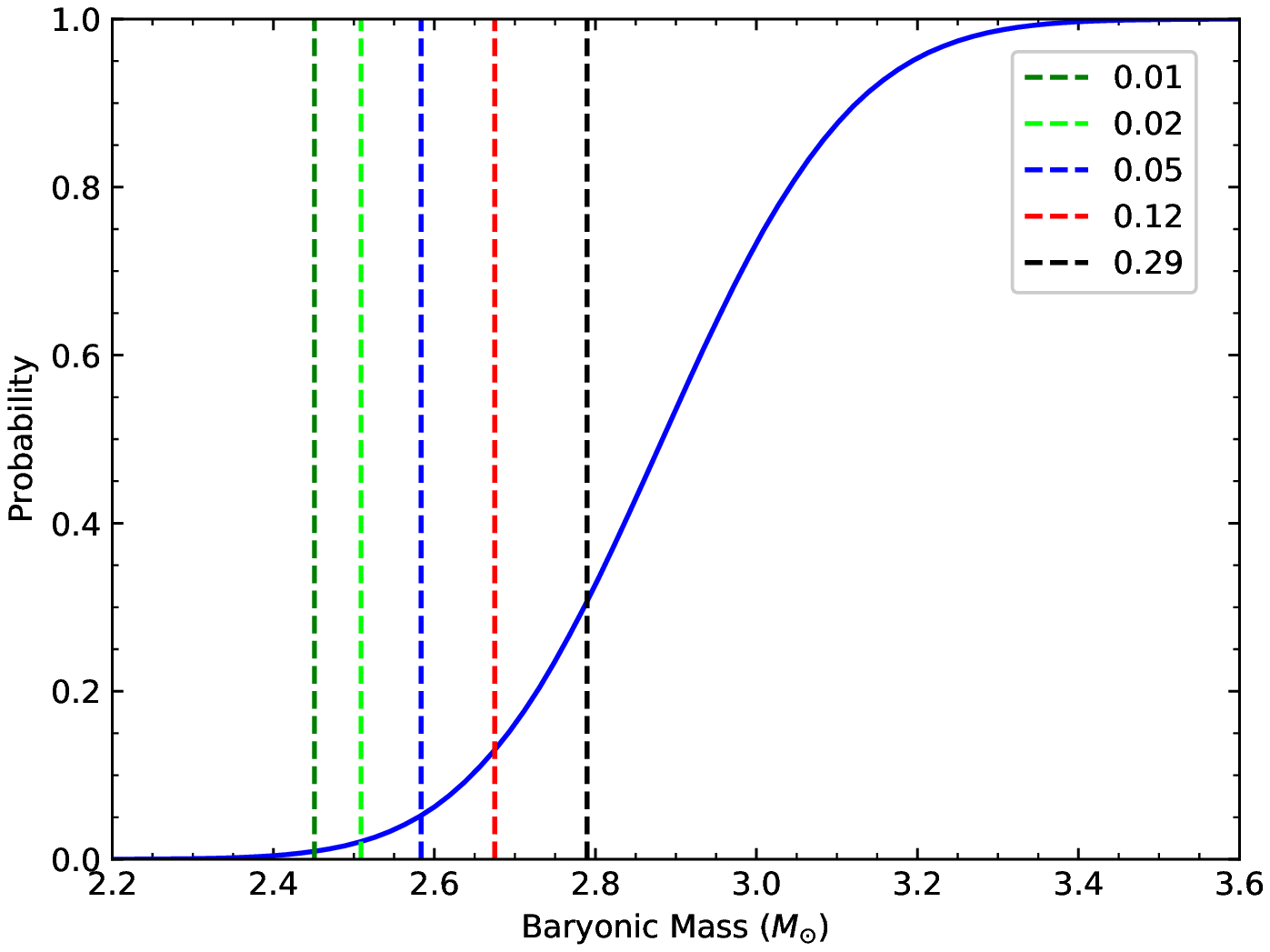}
\includegraphics[width=0.3\textwidth]{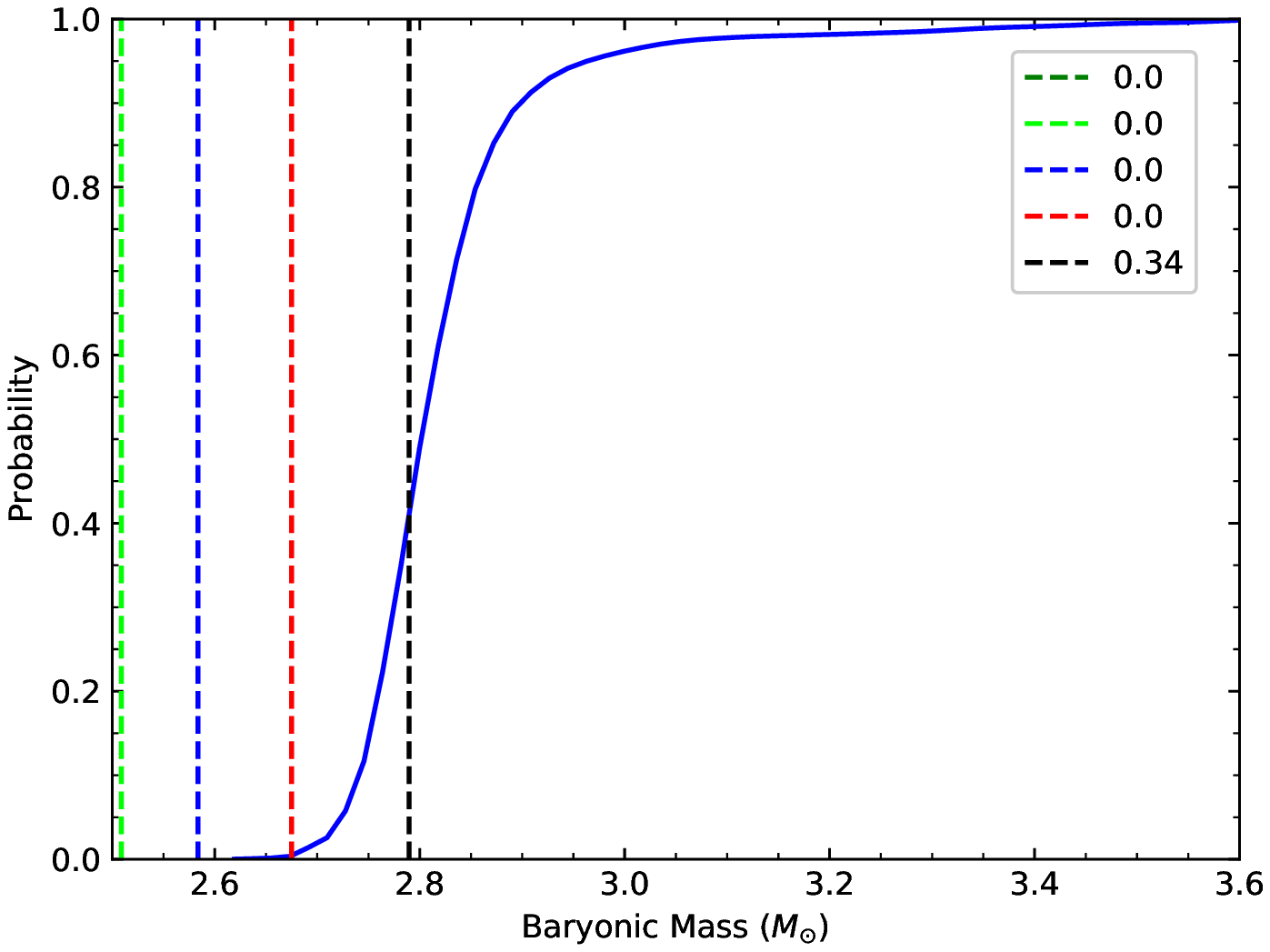}
\includegraphics[width=0.3\textwidth]{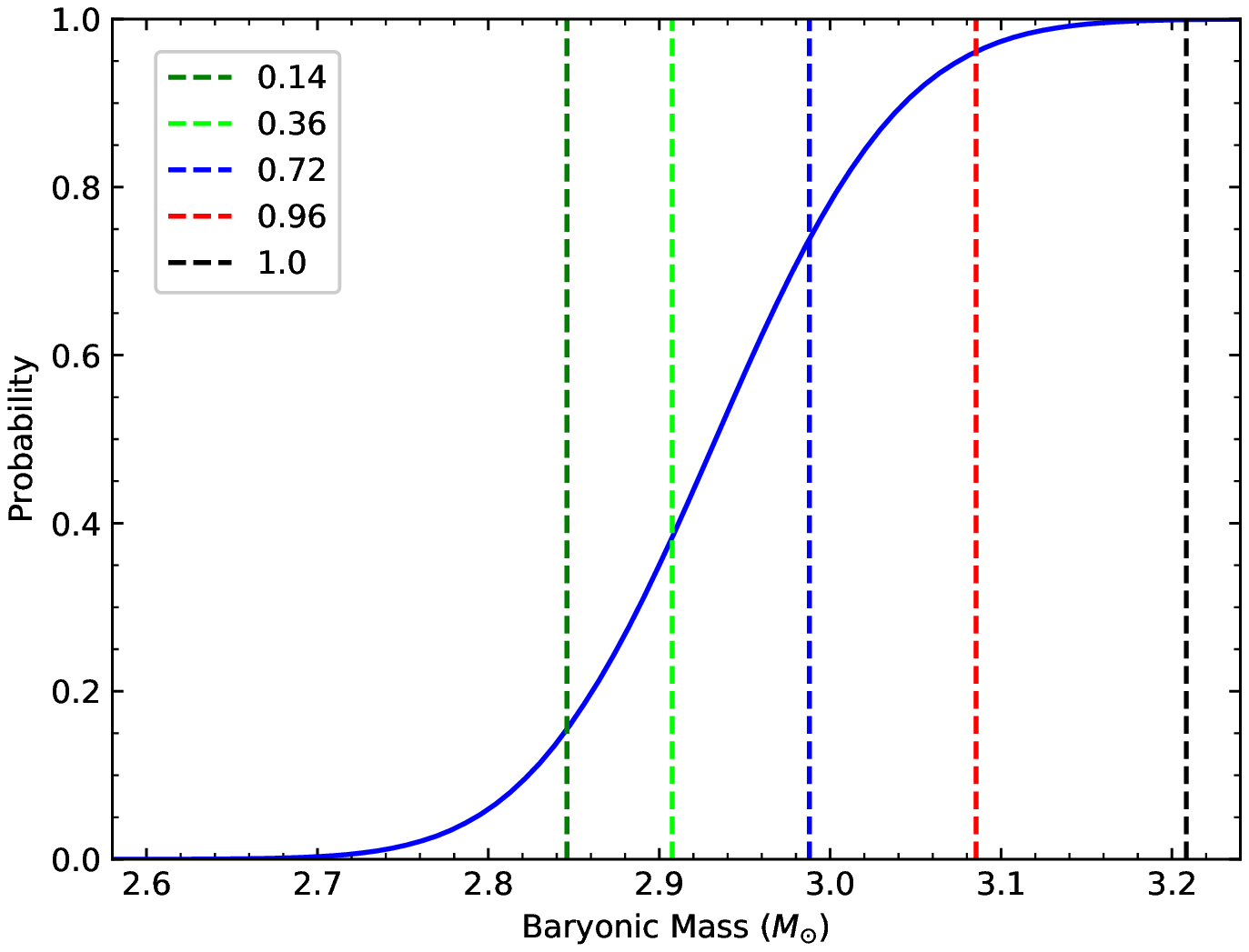}
\includegraphics[width=0.3\textwidth]{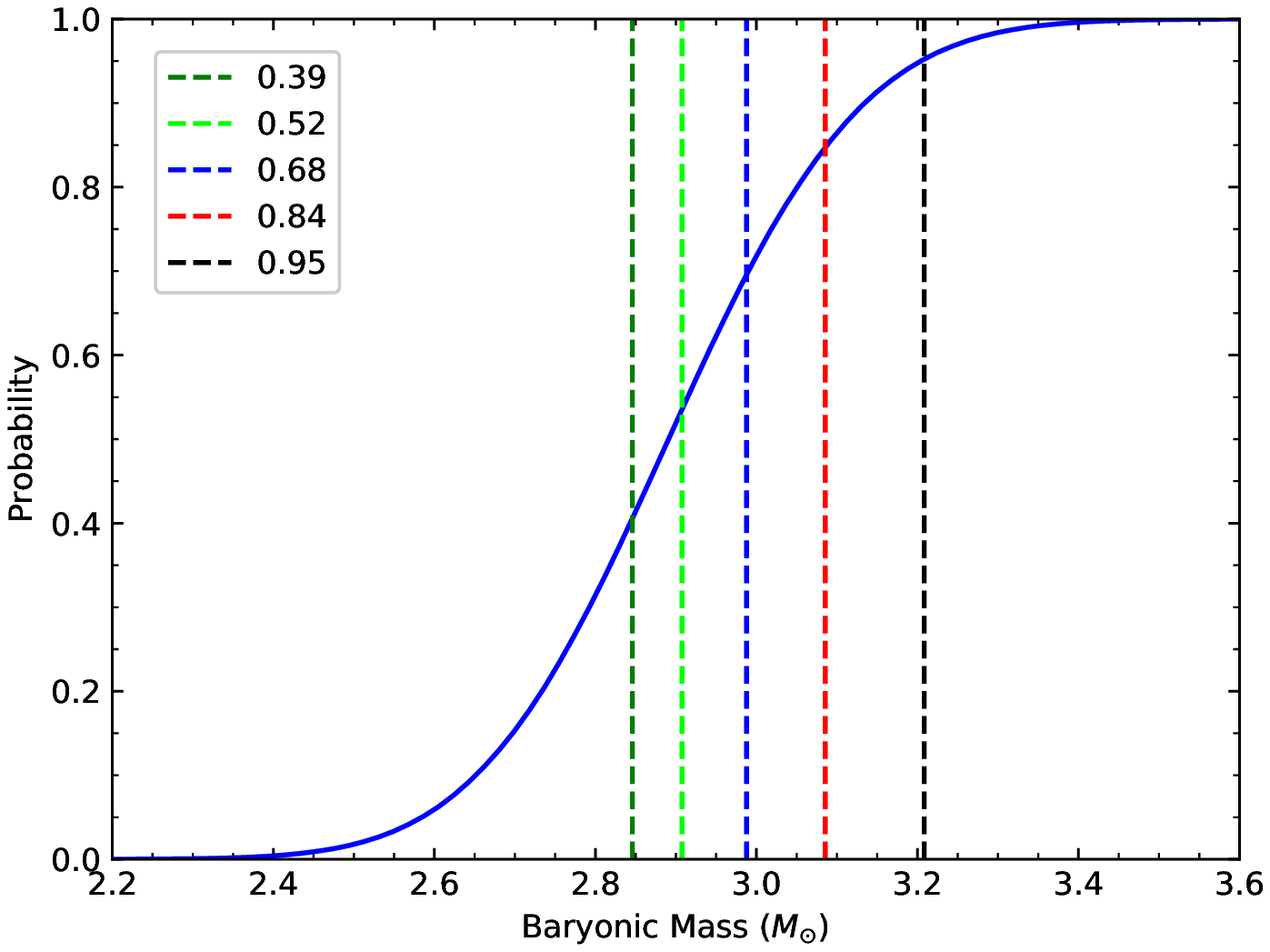}
\includegraphics[width=0.3\textwidth]{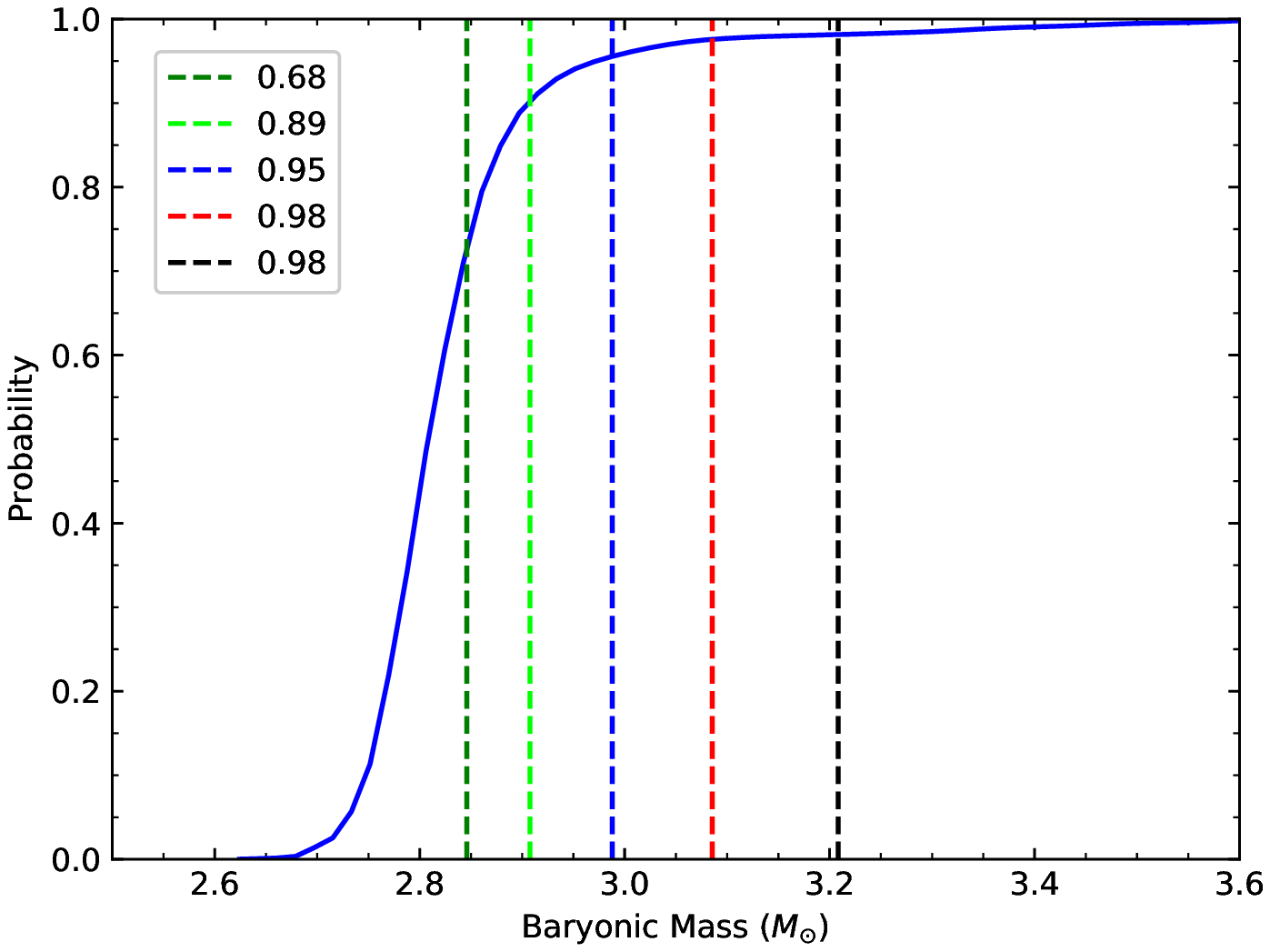}
\caption{{The same as Fig. \ref{fig:simulation-distribution} but for the relatively stiff EoSs (i.e., $R_{1.35M_\odot}\geq 12$ km). Each set of panels from top to bottom is for H4, MPA1, ALF2 and ENG model, respectively}}
\label{fig:simulation-distribution2}
\end{figure}

We have simulated 1 million binary NS mergers for a given EoS model and a given mass distribution scenario. The results are presented in Fig.\ref{fig:simulation-distribution} (for a few soft EoSs yielding $R_{1.35M_\odot}<12$ km) { and Fig.\ref{fig:simulation-distribution2} (for some relatively stiffer EoSs predicting $R_{1.35M_\odot}\geq 12$ km).} The vertical axis represents the possibility of forming a remnant with a baryonic mass smaller than a given value and the horizontal axis represents the remnant baryonic mass. Due to the relatively narrow mass distribution of the pre-merger NSs, the baryonic mass of the remnants have a concentrated distribution, too.
 Thanks to the larger $\sigma$ in the single Gaussian mass distribution scenario, the baryonic mass of the remnants have a wider distribution than that expected in the double Gaussian mass distribution scenario. Therefore, the predicted fractions of forming SMNSs  in these two mass distribution scenarios are significantly different for relatively small $j/j_{\rm Kep}\sim (0.6,~0.7,~0.8)$. For $j/j_{\rm Kep}\sim 1$, { the adopted EoS models but except H4 and ALF2,} predicted formation of SMNSs (and even a small amount of stable NSs) in more than half of the mergers. In particular, all the three  mass distribution models suggest fractions high up to $\sim 90\%$ for the EoSs of APR4, MPA1 and ENG.  Since the data of local short GRBs suggest a NS merger that is likely comparable to the rate inferred from the gravitational wave data \citep{Jin2017}, as for the APR4 model, the data may suggest $j/j_{\rm Kep}\leq 0.9$ supposing the short GRBs can only powered by the black hole central engine. { The uncertainty for such a speculation is whether a gravitational collapse is indeed necessary for powering short GRBs. It is also evident that the mass distribution model adopted from \citet{Fryer} gives the highest SMNS formation rate, since the resulting $M_{\rm tot}$ is statistically smaller than the other two scenarios. In the near future the gravitational wave data will test whether it is the case or not.}

\section{$j/j_{\rm Kep}$: constraints with GW170817 and its role in shaping the bounds on $M_{\rm max}$}
A $M_{\rm tot}=2.74^{+0.04}_{-0.01}~M_\odot$ is inferred for GW170817 \citep{Abbott2017}. In this section we aim to set a ``tight" constraint on $j/j_{\rm Kep}$ of the newly formed remnants in some specific EoS models (in addition to seven models discussed in the pervious section, here we include also the WFF2 and APR3 models that are also allowed by the gravitational wave data). For such a purpose, we assume the double NSs have the same mass (i.e., each has a gravitational mass $M_1=M_2=1.365~M_\odot$), for which usually the total baryonic mass reaches almost the minimum (supposing the radius of NS is insensitive of its mass, with eq.(\ref{eq:BE}) it is straightforward to show that the total BE minimizes for $M_1=M_2$). While for a $M_{\rm tot}>2.73M_\odot$ and $M_1 \neq M_2$, the chance of forming a black hole is enhanced and the upper bound on $j/j_{\rm Kep}$ is less tighter.

\begin{figure}[h]
\centering
\includegraphics[width=0.45\textwidth]{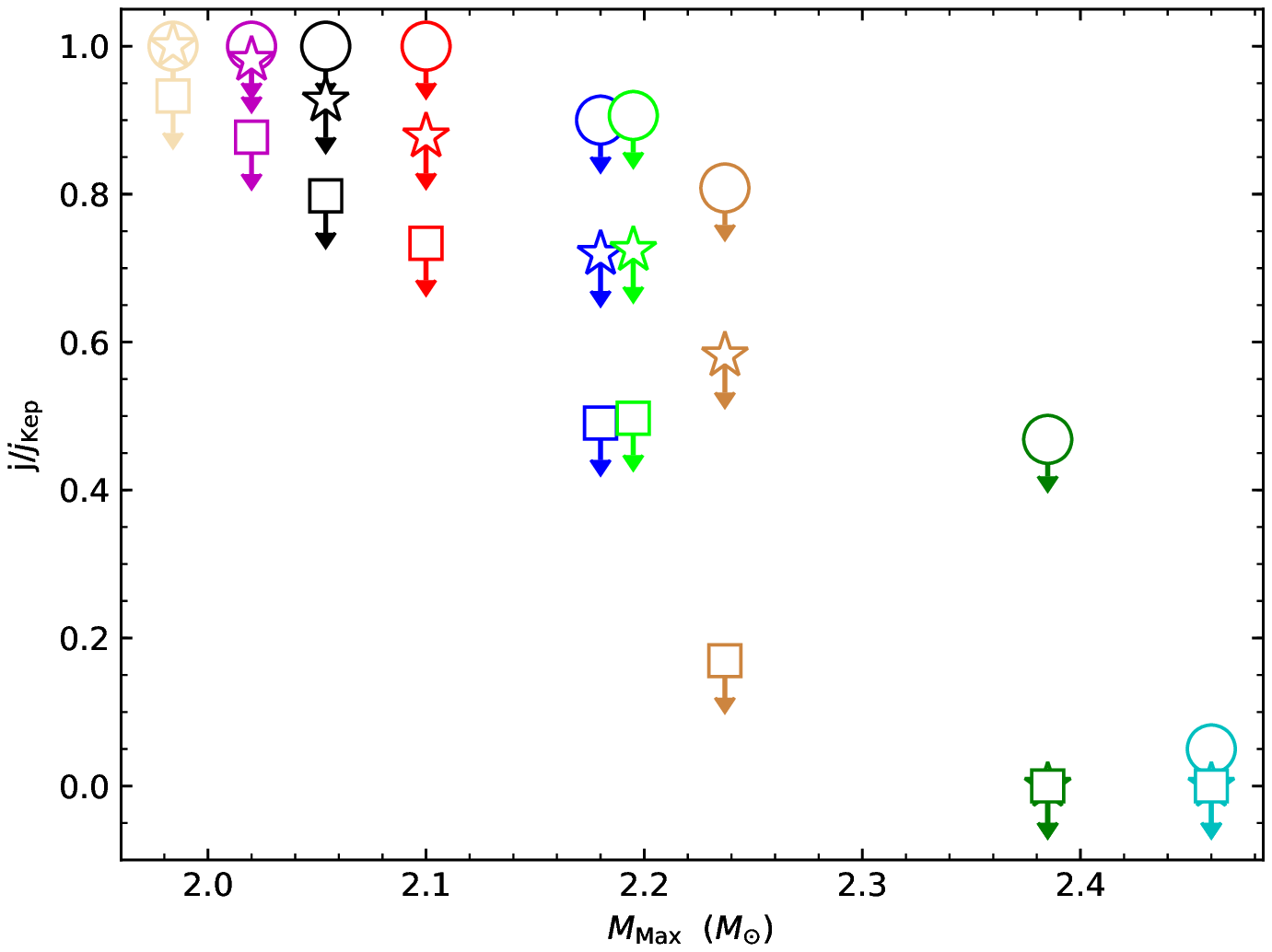}
\includegraphics[width=0.45\textwidth]{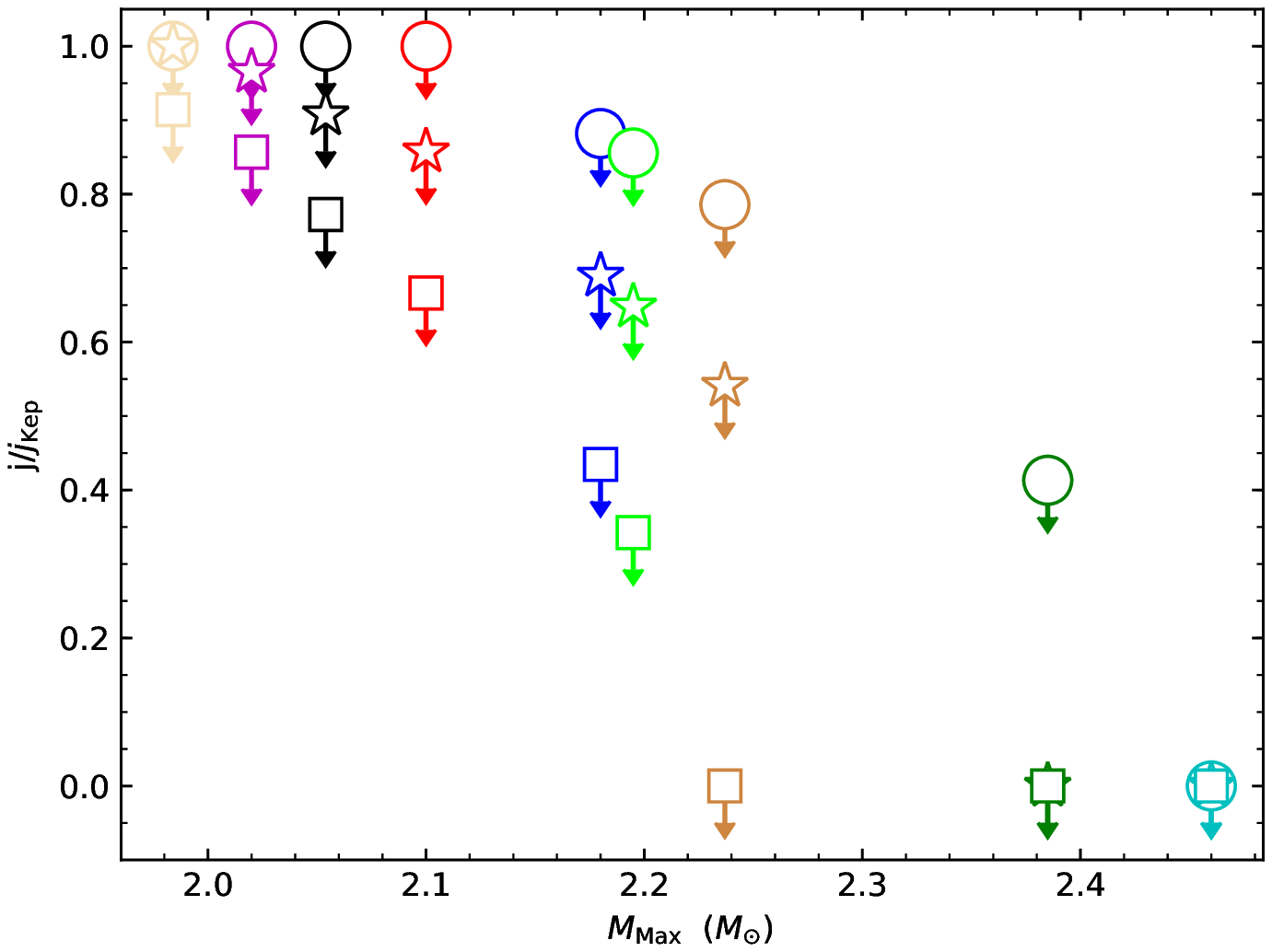}
\caption{The constraint on $j/j_{\rm Kep}$, set by the absence of SMNS signature for GW170817/GRB 170817A, for the ALF2, H4, SLy, NS radius-based EoS, APR4, WFF2, ENG, APR3 and MPA1 models, from left to right in each panel, where $M_{\rm tot}=2.73M_\odot$ and $M_1=M_2$ are assumed (see the circles). The left and right panels are for $m_{\rm loss}=0.03M_\odot$ and $0.05M_\odot$, respectively. The requests on $j/j_{\rm Kep}$, supposing a SMNS is still absent for mergers of $M_{\rm tot}=(2.5,~2.6)M_\odot$ binary systems, are also presented (see the open squares and pentagrams, respectively).} \label{fig:EoS-j}
\end{figure}

The presence of a short GRB (i.e., GRB 170817A) in about two seconds after GW170817 as long as the absence of reliable evidence for a relatively long-lived massive NS in the afterglow emission of GRB 170817A requires that
\begin{equation}
M_1+{\rm BE}_1+M_2+{\rm BE}_2-m_{\rm loss}>M_{\rm b,crit},
\label{eq:constraint-3}
\end{equation}
which can then yield a constraint on $j/j_{\rm Kep}$ if the above condition is violated for $j=j_{\rm Kep}$. Again, we assume $m_{\rm loss}=0.03M_\odot$ and $0.05M_\odot$, respectively. Indeed, for {\it the nine EoS models}  considered in this section we find that ALF2, H4, SLY and the NS-radius based EoS models are well consistent with the data, while for MPA1, APR3, APR4, { WFF2 and ENG} we need $j<(0.05,~0.47,~0.89,~0.91,~0.81)j_{\rm Kep}$, as shown in Fig.\ref{fig:EoS-j}. Therefore, the absence of a SMNS signal in GW170817
{ has ruled out the MPA1 model because the required $j/j_{\rm Kep}$ to form a black hole is too low to be realistic. The APR3 model is also challenged.}

There are five double NS systems in the Galaxy has a $M_{\rm tot}\leq 2.6M_\odot$ \citep{Lattimer2012}, hence it is reasonable to speculate frequent mergers of these relatively ``light" binaries. Let us briefly examine what will happen in such mergers. Two ``symmetric" mergers (i.e., $M_1=M_2$) of the binaries with $M_{\rm tot}=(2.5,~2.6)M_{\rm tot}$, respectively, are investigated. Figure \ref{fig:EoS-j} shows that even for H4, SLy and the NS-radius based EoS models, $j/j_{\rm Kep}<1$ is necessary to form black holes rather than SMNSs if $M_{\rm tot}\leq 2.6M_\odot$ (This conclusion applies to ALF2 only if $M_{\rm tot}\leq 2.5M_\odot$). The NS Interior Composition Explorer \citep[NICER;][]{Gendreau2016}, a space mission dedicated to accurately measure the NS radius, was successfully launched in June 2017. The NS-radius based EoS is thus expected to be further improved in the near future, with which the possibility
of $j=j_{\rm Kep}$ can be directly probed by the gravitational wave data together with the electromagnetic counterpart data. The second generation detectors, like advanced LIGO/Virgo, may be unable to reliably measure the gravitational wave radiation of the pre-collapse massive NSs unless the sources are nearby \citep{Abbott2017b}. An independent way, like that figured out here, is very helpful in revealing the kinetic rotational energy loss of the remnants.

\begin{figure}[h]
\centering
\includegraphics[width=0.5\textwidth]{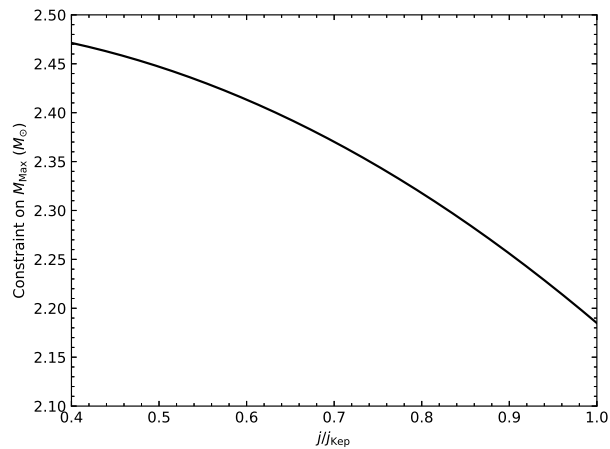}
\caption{The 90\% confidence level upper limits on $M_{\rm max}$, set by the data of GW170817, as a function of $j/j_{\rm Kep}$. The $m_{\rm loss}=0.03M_\odot$ is assumed. For $m_{\rm loss}=0.05M_\odot$ the constraints on $M_{\rm max}$ will be tighter by $\sim 0.02M_\odot$.} \label{fig:Mmax}
\end{figure}

{ With the data of GW170817 and the mass-shedding limit assumption, very tight constraint on $M_{\rm max}<2.17M_\odot$ \citep{Rezzolla2017} has been reported (see also \citet{Margalit2017}  based on the arguments of the limits on the kinetic energy). Though such a bound is well consistent with that of the NS radius based EoS \citep[i.e., $\leq 2.15M_\odot$; see][]{Ozel2016}, here we would like to emphasize the uncertainty caused by the possibility of $j/j_{\rm Kep}<1$. Our approach is as follows. The upper limit of $M_{\rm max}$ could be set by eq.(\ref{eq:constraint-3}) given the mass of each NS, their compactness and the compactness at $M_{\rm max}$. In order to estimate the mass distribution of each NS derived by LIGO's data, we approximate the posterior distribution of chirp mass ${\cal M}=(m_1\times m_2)^{3/5}/(m_1+m_2)^{1/5}$ and mass ratio $q=m_2/m_1$ (note that $q\leq 1$) by normal distribution. Specifically, we have $P({\cal M}, q)\propto \exp{\left[-({\cal M}-\mu_{\cal M})^2/2\sigma_{\cal M}^2 -(q-\mu_q)^2/2\sigma_q^2 \right]}$, where $\mu_{\cal M}=1.188$, $\sigma_{\cal M}=0.002$, $\mu_q=0.85$ and $\sigma_q=0.09$ \citep[see also][however, they took $\mu_q=1$, with which the result is slightly different from ours]{{Margalit2017}}. Then we properly convert the distribution $P({\cal M}, q)$ to the distribution $P(m_1, m_2)$ by Jacobian Matrix. The compactness relies on the radii of each NS and we fix the radius of them to the radius at $M_{\rm max}$ in order to give the upper limit of their compactness. Note that the radius at $M_{\rm max}$ is the lower limit of the NS radius. According to \citet{Ozel2016}, the NS with a mass of $0.8-1.8~M_\odot$ are likely within the range of $9.6~{\rm km}<R<11.4~{\rm km}$ {(note that a lower limit of $\sim 10.7$ km for a non-rotating NS with a mass of $1.6~M_\odot$  is suggested in \citet{Bauswein2017} based on the data of GW170817)}. The causality also gives a theoretical constraint on the lower limit of NS radius\citep{Lattimer2012}. We generate the mass samples by Monte Carlo simulation, and implement the two independent constraints (i.e. the causality constraint and observational constraint) of NS radius to calculate the corresponding distribution of the upper limit of $M_{\rm max}$ for different $j/j_{\rm Kep}$. Our calculation shows that the result is insensitive to the radius and the two results are almost indistinguishable. The $90\%$ confidence level upper limits on $M_{\rm max}$, as a function of $j/j_{\rm Kep}$, are presented in Fig.\ref{fig:Mmax}. For $j=j_{\rm Kep}$ and $m_{\rm loss}=0.03M_\odot$ we have $M_{\rm max}<2.19M_\odot$, consistent with \citet{Margalit2017} and \citet{Rezzolla2017}. }

\section{Summary}
The gravitational wave data of GW170817 strongly favor the equation of state models that predict compact neutron stars, which is well in support of the
spectroscopic radius measurement results of a group of neutron stars obtained during
thermonuclear bursts or in quiescence \citep{Lattimer2016,Ozel2016}. Motivated by such a remarkable progress and benefited from the
discovery of a universal relation between the maximum mass of a NS and the normalized angular momentum \citep{Beru2016}, in this work we have examined the fate of the NS merger remnants and focus on the roles of the angular momentum and the mass distribution of the binary NSs.
Some adopted EoS models (including SLy, APR4, the NS radius based EoS, ALF2, ENG and MPA1) yield SMNSs in more than half of the mergers for $j=j_{\rm Kep}$ (i.e., the so-called mass shedding limit; see Fig.\ref{fig:simulation-distribution} and Fig.\ref{fig:simulation-distribution2}). While for $j\lesssim 0.7j_{\rm Kep}$, the possibility of producing a non-negligible fraction of SMNSs formed in the mergers depends sensitively on both the EoS and the mass distribution of the binary systems. For the current unique event GW170817/AT2017gfo, the absence of a SMNS signature already rules out the MPA1 model and possibly also the APR3 model.  In the next decade, hundreds of NS merger events will be recorded by the second generation gravitational wave detectors, as implied by the successful detection of GW170817 in the O2 run of advanced LIGO and also by the local short GRB data. The statistical study of such a large sample will better determine the mass distribution function of the pre-merger NSs. Moreover, the statistical study of these events, in particular those with a relatively small $M_{\rm tot}\leq 2.6M_\odot$, would shed valuable light on possible significant rotational kinetic energy loss via high frequency gravitational waves and thermal neutrinos (see Fig.\ref{fig:EoS-j}). NICER was launched in June 2017 to accurately measure the NS radius, hence the NS-radius based EoS is expected to be further improved in the near future. With a better determined EoS model, the possibility that the uniform rotations of pre-collapse remnants formed in binary NS mergers reach the mass-shedding limit can be indirectly probed by the gravitational wave data together with the electromagnetic counterpart data. { For GW170817-like or even closer sources, the advanced LIGO/Virgo in their full sensitivity run may be able to directly detect the gravitational wave radiation of the pre-collapse massive NSs and thus provide further insight into such scenarios.}
Finally, we would like to remark that if the uniform rotation of the pre-collapse remnant does not reach the mass shedding limit,  the constraint on $M_{\rm max}$ could be substantially loosened (see Fig.\ref{fig:Mmax}).

\section*{Acknowledgments}
We thank the anonymous referee for the helpful suggestions. This work was supported in part by 973 Programme of China (No. 2014CB845800), by NSFC under grants 11525313 (the National Natural Fund for Distinguished Young Scholars), 11273063 and 11433009, by the Chinese Academy of Sciences via the Strategic Priority Research Program (No. XDB09000000) and the External Cooperation Program of BIC (No. 114332KYSB20160007).

\clearpage

\end{document}